\documentclass{ws-rv9x6}
\usepackage{subfigure}   % required only when side-by-side / subfigures are used
\usepackage{ws-rv-thm}   % comment this line when `amsthm / theorem / ntheorem` package is used
\usepackage{ws-rv-van}   % numbered citation & references (default)
\usepackage[greek,english]{babel}
\usepackage{teubner}
\makeindex
\RequirePackage{color}
\definecolor{MyDarkGreen}{rgb}{0.02,0.60,0.06}

\renewcommand{\theequation}{\thesection.\arabic{equation}}
\DeclareFontFamily{OT1}{pzc}{}
\DeclareFontShape{OT1}{pzc}{m}{it}{<-> s * [0.900] pzcmi7t}{}
\DeclareMathAlphabet{\mathpzc}{OT1}{pzc}{m}{it}
\def\q{\hbox{\foreignlanguage{greek}{\coppa}}}
\def\qq{{\hbox{\foreignlanguage{greek}{\footnotesize\coppa}}}}
\def\oureta{{{\eta_Q}}}
\def\ourG{{G_Q}}
\def\ourD{{D_Q}}
\def\ouretahat{{\hat{{\eta}}_Q}}

\begin{document}

\chapter[Scaling and Finite-Size Scaling above the Upper Critical Dimension]{Scaling and Finite-Size Scaling above the Upper Critical Dimension}
\label{ra_ch1}

\author[R. Kenna and B. Berche]{Ralph Kenna}%\footnote{Author footnote.}}
%\index[aindx]{Author, F.} % or \aindx{Author, F.}
%\index[aindx]{Author, S.} % or \aindx{Author, S.}
\address{$^1$ Applied Mathematics Research Centre, Coventry University, \\
Coventry CV1 5FB, England,\\
r.kenna@coventry.ac.uk}

\author[R. Kenna and B. Berche]{Bertrand Berche}%\footnote{Author footnote.}}
%\index[aindx]{Author, F.} % or \aindx{Author, F.}
%\index[aindx]{Author, S.} % or \aindx{Author, S.}

\address{$^2$Statistical Physics Group, Institut Jean Lamour,
UMR CNRS 7198,\\ Universit{\'{e}} de Lorraine
B.P. 70239, 54506 Vand{\oe}uvre l{\`{e}}s Nancy Cedex, France\\
bertrand.berche@univ-lorraine.fr\\
{\rm Original date: October 15th, 2014}}

\begin{abstract}
In the 1960's, four famous scaling relations were developed which relate the six standard critical exponents describing continuous phase transitions in the thermodynamic limit of statistical physics models. 
They are well understood at a fundamental level through the renormalization group.  
They have been verified in multitudes of theoretical, computational and experimental studies
and are firmly established and profoundly important for our understanding of critical phenomena.  
One of the scaling relations, hyperscaling, fails above the upper critical dimension. 
There, critical phenomena are governed by Gaussian fixed points in the renormalization-group formalism. Dangerous irrelevant variables are required to deliver the mean-field and Landau values of the critical exponents, which are deemed valid by the Ginzburg criterion. 
Also above the upper critical dimension, the standard picture is that, unlike for low-dimensional systems, finite-size scaling is non-universal.
Here we report on new developments which indicate that the current paradigm is flawed and incomplete. 
In particular, the introduction of a new exponent characterising the finite-size correlation length allows one to extend hyperscaling beyond the upper critical dimension. 
Moreover, finite-size scaling is shown to be universal provided the correct scaling window is chosen. 
These recent developments also lead to the introduction of a new scaling relation analogous to one introduced by Fisher 50 years ago.
\end{abstract}
%\markright{Customized Running Head for Odd Page} % default is chapter title.
\body

%%%%%%%%%%%%%%%%%%%%%%%%%%%%%%%%%%%%%%%%%%%%%%%%%%%%%%%%%%%%%%%%%%%%%%%%%%%
\newpage
%\mycheck{\red{Introduction}}
\section{Introduction}
\label{Introduction}
\setcounter{equation}{0}
%%%%%%%%%%%%%%%%%%%%%%%%%%%%%%%%%%%%%%%%%%%%%%%%%%%%%%%%%%%%%%%%%%%%%%%%%%%

In the standard picture, critical phenomena above the upper critical dimension are believed to be well described by mean-field theory and by Landau theory. 
This picture is supported by the renormalization-group formalism coupled to Fisher's dangerous irrelevant variables concept.
In the conventional paradigm, hyperscaling fails above the upper critical dimension and finite-size scaling ceases to be universal. 
%This is the region inaccessible to experiment for short-range statistical physics models and trivial in the quantum field theory sense. 
%Indeed, i
In many reviews and textbooks, theories with short-range interactions beyond the upper critical dimension play a precursory role, appearing in the early, introductory, chapters, where they are seen as a step en route to more physically interesting phenomena at or below upper dimensionality.
For this  reason, high-dimensional  theories play a foundational role in statistical mechanics. 
Increasing the interaction range reduces the upper critical dimension.
Such experimentally accessible models may be of direct physical relevance. 
Over the last  decades, however, it has been recognised that the current paradigm is not satisfactory. 
We quote from Ref.~[\refcite{Binderquote}] in the context of  $\phi^4$ theory:
\begin{quote}
``Thus we arrive at a rather disappointing state of affairs - although
for the $\phi^4$ theory in $d=5$ dimensions all exponents are known, including
those of the corrections to scaling, and in principle very complete
analytical calculations are possible, the existing theories clearly are
not so good.''
\end{quote}

Very recently, progress has been made to provide a complete account of scaling above the upper critical dimension. 
Given the basic role mean-field and Landau theories play, this development is important in a fundamental sense. 
The  aims of this exposition are to contextualise and present the recent theoretical advances.
We begin in Section~\ref{II} with a summary of scaling and finite-size scaling at continuous phase transitions, including of the scaling relations. 
Derived in the 1960's, these form a cornerstone for the statistical physics of phase transitions. 
We also touch upon logarithmic corrections to scaling, a subject reviewed in a previous volume of this series \cite{KeVol3}.
In Section~\ref{11.05.2009c} we revisit Widom's scaling ansatz and Kadanoff's block-spin renormalization.
Mean-field theory and Landau theory are developed as a foundation for $\phi^4$ theory in Section~\ref{III}, theory, where the Ginzburg criterion is recalled.
In Section~\ref{secwilsonRG}, Wilson's renormalization-group formalism for the Gaussian fixed point is briefly outlined.
To describe scaling above the upper critical dimension, dangerous irrelevant variables have to be accounted for, and these are treated in both infinite and finite systems in Section~\ref{DIV}.
Reasons why this standard standard paradigm is ad hoc and unsatisfactory are given in Section~\ref{shortcomings}.
Recent developments are summarised in Section~\ref{longcomings}, where two new exponents $\q$ (pronounced ``koppa'') and $\eta_Q$ are introduced. The new, full scaling and finite-size scaling theory above the upper critical dimension is outlined in Section~\ref{fullthy}. We also include new results for logarithmic corrections {\emph{at}} the critical dimension itself, supplementing the review in Ref.~[\refcite{KeVol3}].
We conclude in Section~\ref{Conclusions}.

%%%%%%%%%%%%%%%%%%%%%%%%%%%%%%%%%%%%%%%%%%%%%%%%%%%%%%%%%%%%%%%%%%%%%%%%%%%
%\mycheck{\red{II}}
\section{Scaling at Continuous Phase Transitions}
\label{II}
\setcounter{equation}{0}
%%%%%%%%%%%%%%%%%%%%%%%%%%%%%%%%%%%%%%%%%%%%%%%%%%%%%%%%%%%%%%%%%%%%%%%%%%%
\index{phase transition}  

In this section, we briefly summarise the standard power-law scaling paradigm at continuous phase transitions and the relations between the various critical exponents. 
This includes a description of standard finite-size scaling (FSS). 
We also very briefly summarise the new developments which are elucidated in subsequent sections.

Although second-order or continuous phase transitions
also appear in particle physics, cosmology, fluid mechanics,  and other areas of physics, 
we employ the language of magnetism in this exposition for purposes of clarity. 
It is straightforward to convert the results to the terminology of related fields.

To begin this section, we introduce the basic functions which describe global and local properties of the system. These are called thermodynamic and correlation functions respectively. 
Since the latter concern system details at a microscopic level, their definitions are model specific and throughout this work, we use the Ising and $\phi^4$ models as generic representations of spin models and field theory.

%%%%%%%%%%%%%%%%%%%%%%%%%%%%%%%%%%%%%%%%%%%%%%%%%%%%%%%%%%%%%%%%%%%%%%%%%%%
%\mycheck{\red{Thermodynamic functions}}
\subsection{Thermodynamic functions}
%%%%%%%%%%%%%%%%%%%%%%%%%%%%%%%%%%%%%%%%%%%%%%%%%%%%%%%%%%%%%%%%%%%%%%%%%%%
\index{thermodynamic function}

We consider a system of spins $s_i$ located at the sites $i$ of a $d$-dimensional lattice with $L$ sites in each direction, so that there are  $N=L^d$ sites overall.
If the lattice constant is $a$, the volume of the system is $Na^d$.
The partition function\index{partition function} is defined as
%
%\mycheck{\red{II1}}
\begin{equation}
 Z_L = \sum_{\{s_i\}}{e^{-\beta \mathcal{H}}},
 \label{II1}
\end{equation}
where $\mathcal{H}$ is the total energy associated with a given  configuration $\{s_i\}$ of the spins and $\beta = 1/k_BT$ is the reciprocal of the Boltzmann constant $k_B$ times the absolute temperature $T$.
The Hamiltonian is 
%
%\mycheck{\red{II2}}
\begin{equation}
 {\mathcal{H}} = E - H M,
 \label{II2}
\end{equation}
in which $E$ represents the energy due to interactions between the spins themselves, $H$ is the strength of an external magnetic field and $M$ is the magnetisation of a given configuration. 
For later convenience we define the reduced external field as $h = \beta H$. 

The Ising model is defined through the configurational energy and magnetisation\cite{Ising,Lenz}
%\mycheck{\red{II10,11}}
\begin{eqnarray}
 E & = & - J \sum_{\langle{ij}\rangle}{s_is_j},
 \label{II10}
\\
 M & = & \sum_i{s_i} ,
 \label{II11}
\end{eqnarray}
in which $s_i \in \{{\pm 1}\}$.
The Helmholtz free energy\index{energy! Helmholtz free} is defined as
%\mycheck{\red{II3}}
\begin{equation}
 F_L=-k_B T \ln{Z_L}.
 \label{II3}
\end{equation}
We denote intensive quantities in lower case, so that, e.g., 
%
%\mycheck{\red{II4}}
\begin{equation}
 f_L= \frac{F_L}{N}.
 \label{II4}
\end{equation}
At $h=0$, the entropy is given by
%
%\mycheck{\red{II5}}
\begin{equation}
 Ns_L = -\frac{\partial F_L}{\partial T} = k_B \ln{Z_L} + \frac{1}{T}Ne_L,
 \label{II5}
\end{equation}
where
%
%\mycheck{\red{II6}}
\begin{equation}
 Ne_L= -\frac{\partial \ln{Z_L}}{\partial \beta} = \langle{E}\rangle,
 \label{II6}
\end{equation}
is the internal energy and $\langle{\dots}\rangle$ refer to expectation values.
The specific heat is 
%
%\mycheck{\red{II7}}
\begin{equation}
 c  =  \frac{\partial e}{\partial T} 
        = \frac{k \beta^2}{N} \left(\langle{E^2}\rangle - \langle{E}\rangle^2 \right).
\label{II7}
\end{equation}
The magnetisation and magnetic susceptibility are respectively defined  as
%
%\mycheck{\red{II8,II9}}
\begin{eqnarray}
 m & = & - \frac{\partial f }{\partial H } 
     =  \frac{\langle M \rangle}{N} ,
 \label{II8}
\\
 \chi & = & \frac{\partial m}{\partial H} 
        = \frac{ \beta}{N} \left( \langle{M^2}\rangle - \langle{M}\rangle^2 \right).
 \label{II9}
\end{eqnarray}

%%%%%%%%%%%%%%%%%%%%%%%%%%%%%%%%%%%%%%%%%%%%%%%%%%%%%%%%%%%%%%%%%%%%%%%%%%%
%\mycheck{\red{Correlation functions}}
\subsection{Correlation functions}
%%%%%%%%%%%%%%%%%%%%%%%%%%%%%%%%%%%%%%%%%%%%%%%%%%%%%%%%%%%%%%%%%%%%%%%%%%%
\index{correlation function}

For the Ising model, the magnetisation defined in Eq.(\ref{II8}) is 
%
%\mycheck{\red{II12}}
\begin{equation}
 m = \frac{1}{N} \sum_i{\langle{s_i}\rangle}.
\label{II12}
\end{equation}
In the translation-invariant case where  $\langle{s_i}\rangle$ is independent of $i$, this simplifies to $ m = \langle{s_i}\rangle$.
This may be expected when the system is infinite in extent.
There is no spontaneous magnetisation in a finite-size system.
For the spin model, the connected correlation function is defined as
%
%\mycheck{\red{LB1.4.1}}
\begin{equation}
 G(r_i,r_j) = \langle{s_i s_j}\rangle - \langle{s_i}\rangle\langle{s_j}\rangle,
 \label{LB1.4.1}
\end{equation}
where $r_i$ represents the position of the $i$th lattice site.
%while the first term alone may be refered to as the correlation function.
It is useful to promote $h$ to a function of position, so that the external magnetic field
is not uniform.
One can easily show that the connected correlation function is 
%
%\mycheck{\red{II14}}
\begin{equation}
G(r_i,r_j) = 
  - \frac{\beta}{N} \frac{\partial^2 f}{\partial h_i \partial h_j},
\label{II14}
\end{equation}
where $h_i$ is the reduced field at site $i$.
%Because the influence of a given (local) spin cannot extend indefinitely (globally), the connected correlation function should vanish in the large-distance limit.
%This feature is known as  the {\emph{clustering property}}.

%%%%%%%%%%%%%%%%%%%%%%%%%%%%%%%%%%%%%%%%%%%%%%%%%%%%%%%%%%%%%%%%%%%%%%%%%%%
%\mycheck{\red{plcps}}
\subsection{Power-law critical-point scaling}
\label{plcps}
%%%%%%%%%%%%%%%%%%%%%%%%%%%%%%%%%%%%%%%%%%%%%%%%%%%%%%%%%%%%%%%%%%%%%%%%%%%

We are interested in behaviour near the critical point $(T,H)=(T_c,0)$.
To gain a dimensionless measure of distance from criticality we define the reduced temperature
%\mycheck{\red{II15}}
\begin{equation}
 t= \frac{T-T_c}{T_c}. 
\label{II15}
\end{equation}
We henceforth write thermodynamic functions in terms of reduced variables, e.g.,   $c_L(t,h)$, $m_L(t,h)$ and $\chi_L (t,h)$. 
To make the connection to the macroscopic thermodynamical world, statistical mechanics is taken to its large $L$ limit.
There the leading power-law scaling behaviour\index{scaling! leading, power-law} which captures the dependencies of these thermodynamic functions at a phase transition of second order is given by
%\mycheck{\red{e,c,mt,chi}}
\begin{eqnarray}
  c_\infty(t,0) \sim |t|^{-\alpha} ,  \label{c}
\\
 m_\infty(t,0) \sim (-t)^{\beta}   ,  \label{mt}
\\
 m_\infty(0,h) \sim h^{\frac{1}{\delta}}   ,  \label{mh}
\\
 \chi_\infty(t,0) \sim |t|^{-\gamma} ,
 \label{chi}
\end{eqnarray}
which define the critical exponents\index{critical exponent} $\alpha$, $\beta$, $\gamma$ and $\delta$.
Eq.(\ref{mt}) holds only when $T<T_c$ as there is no spontaneous magnetisation when $T>T_c$.

The  thermodynamic functions\index{thermodynamic function} listed above are derivable from the partition function through Eqs.(\ref{II6})-(\ref{II9}). 
They give the global response of the {\emph{entire}} system to tuning the temperature and/or external field near the phase transition. 
One is also interested in local responses  given by the correlation functions (\ref{LB1.4.1}). 
For systems with translational invariance one  expects  $G(r_i,r_j)$ to depend only on the distance between sites $i$ and $j$ rather than their absolute coordinates.
We write\index{correlation function}
%
%\mycheck{\red{G}}
\begin{equation}
 G_\infty (t,h;r) \sim r^{-p} D \left[{\frac{r}{\xi_\infty(t,h)}}\right],
 \label{G}
\end{equation}
in which $\xi$ represents the correlation length, which measures the  scale of fluctuations away from the fully aligned or fully random states.
The correlation length also diverges close to criticality, and\index{correlation length}
%
%\mycheck{\red{xi}}
\begin{equation}
\xi_\infty(t,0) \sim |t|^{-\nu} , \quad  \quad  \xi_\infty(0,h) \sim {|h|}^{-\nu_c} .
 \label{xi}
\end{equation}
%
%(One may also introduce $\nu_c$ and consider the dependency of correlation length on field as $\xi_\infty(0,h) \sim {h}^{-\nu_c}$.)
When $r \ll \xi_\infty(t)$,  the power-law dominates Eq.(\ref{G}) and one writes
%
%\mycheck{\red{G1}}
\begin{equation}
 G_\infty (t,0;r) \sim r^{-(d-2+\eta)} ,
 \label{G1}
\end{equation}
in which  $\eta$ is  the anomalous dimension.\index{anomalous dimension}

From experiments and Landau theory (Section~\ref{III}) it is expected that the correlation function  decays exponentially away from criticality for sufficiently large distance.
\index{correlation function}
Then $D(y) \sim \exp{(-y)}$, or
%
%\mycheck{\red{G2}}
\begin{equation}
 G_\infty (t,0;r) \sim r^{-p} \exp \left[{-\frac{r}{\xi_\infty(t,0)}}\right],
 \label{G2}
\end{equation}
when $r \gg \xi_\infty(t)$.

The momentum-space equivalent of the general form (\ref{G}) is
%\mycheck{\red{Gmon}}
\begin{equation}
 \tilde{G} (t,h;k)  =   \int_0^\infty{d^d{\mathbf{r}} \ \!G(r)e^{ikr}} 
  =  k^{p-d}g(k\xi),
\label{Gmom}
\end{equation}
for some function  $g$. 
In the critical region, where $p=d-2+\eta$, this becomes
%\mycheck{\red{G1mon}}
\begin{equation}
 \tilde{G} (t,0;k)  \sim    k^{-2+\eta}.
\label{G1mon}
\end{equation}

%%%%%%%%%%%%%%%%%%%%%%%%%%%%%%%%%%%%%%%%%%%%%%%%%%%%%%%%%%%%%%%%%%%%%%%%%%%
%\mycheck{\red{LYfund}}
\subsection{Fundamental theory of phase transitions}
\label{LYfund}
%%%%%%%%%%%%%%%%%%%%%%%%%%%%%%%%%%%%%%%%%%%%%%%%%%%%%%%%%%%%%%%%%%%%%%%%%%%
\index{phase transitions}

In the 1950's, inspired by the fundamental theorem of algebra, Yang and Lee \cite{LY}
developed a {\emph{fundamental theory of phase transitions}}.\index{zeros!Lee-Yang} 
For the finite $L$ Ising model, for example, the partition function in Eq.(\ref{II1}) has a discrete set of zeros in the complex-$h$ plane.
In the limit of infinite volume these Lee-Yang zeros\index{zeros!Lee-Yang}
\index{Lee-Yang zeros}   condense onto curves. In fact, in many circumstances the Lee-Yang theorem ensures that these zeros are purely imaginary\cite{LY}.

When $T>T_c$, the point in the distribution of Lee-Yang zeros closest to the real axis is called the  Yang-Lee edge and we denote it by $h_{\rm{YL}}(t)$. 
It approaches the real  axis as $t$ reduces to its critical value $t=0$ and that approach is also characterised by a power law if the transition is second order, namely
%
%\mycheck{\red{YL}}
\begin{equation}
 h_{\rm{YL}}(t)  \sim t^{\Delta}  ,
 \label{YL}
\end{equation}
in which $\Delta$ is called the gap exponent.

Fisher pointed out that the finite-size partition function \index{partition function} in Eq.(\ref{II1}) also has zeros in the plane of complex temperature. 
In the thermodynamic limit these Fisher\index{zeros!Fisher} 
 zeros\cite{Fi65}  pinch the real temperature axis at the point where the phase transition occurs (namely at $T=T_c$).

%%%%%%%%%%%%%%%%%%%%%%%%%%%%%%%%%%%%%%%%%%%%%%%%%%%%%%%%%%%%%%%%%%%%%%%%%%%
%\mycheck{\red{fsssr}}
\subsection{Finite-size scaling, shifting and rounding}
\label{fsssr}
%%%%%%%%%%%%%%%%%%%%%%%%%%%%%%%%%%%%%%%%%%%%%%%%%%%%%%%%%%%%%%%%%%%%%%%%%%%
\index{finite-size scaling}
\index{shifting}
\index{rounding}

Genuine phase transitions can only appear in systems of infinite size. 
In finite-size systems, the divergences which are characteristic of some functions are replaced by peaks of finite height. 
These peaks have a ``rounded'' structure and  are shifted away from the critical point of the infinite-volume system to what is called the pseudocritical point or effective critical point.

The FSS hypothesis is that the relationship between functions in the thermodynamic limit and their finite-size counterparts enters through the ratio of the two  length scales $L$ and $\xi_\infty$
\cite{FiBe72,Br82,FSS,Barber}. 
For a generic function $P(t,h)$, say,  this relationship is expressed as (setting $h=0$) 
%\mycheck{\red{II31}}
\begin{equation}
 \frac{P_L(t_L)}{P_\infty(t)}
 =
 {\mathcal{F}}_P
 \left[{\frac{L}{\xi_\infty(t)}}\right].
 \label{II31}
\end{equation}
Here, $t_L$ is the pseudocritical value of the reduced temperature -- that at which the function $P_L$    has an extremum.
Here, and henceforth, we have suppressed the dependency of $P$ on $h$ when the latter vanishes.

We suppose $P_\infty(t)\sim |t|^{-\rho}$ in the thermodynamic limit. 
 Fixing the scaling ratio $w = L/\xi_\infty(t)$ in Eq.(\ref{II31}) amounts to the substitution $|t| \rightarrow w^{1/\nu}L^{-1/\nu}$, from which
%\mycheck{\red{II32}}
\begin{equation}
 P_L(t_L)
 =
 {\mathcal{F}}_P (w)
 P_\infty \left({w^{\frac{1}{\nu}} L^{-\frac{1}{\nu}}}\right)
  \sim L^{\frac{\rho}{\nu}}.
 \label{II32}
\end{equation}
In fact the scaling form $P_L(t) \sim L^{\rho / \nu}$ 
usually holds as a good approximation in a region around the peak $t_L$, known as the {\emph{scaling window}}. In many instances the scaling window includes {\emph{both}} the critical point $t=0$ as well as the pseudocritical point $t=t_L$.
In other words, to obtain the FSS behaviour of $P_L$ from $P_\infty(t)$, one simply replaces the infinite-volume correlation length $\xi_\infty(t)$ by the actual length of the system $L$ inside the scaling window. 
Applying it to the functions in Eqs.(\ref{c}), (\ref{mt}), (\ref{chi}), (\ref{xi}), (\ref{G2}), (\ref{YL}), one obtains
%
%\mycheck{\red{cL,mL,chiL,xiL,G2L,YLL}}
\begin{eqnarray}
c_L(t_L) & \sim & L^{\frac{\alpha}{\nu}},
\label{cL} \\
m_L(t_L) & \sim & L^{\frac{-\beta}{\nu}} , 
\label{mL} \\
\chi_L(t_L) & \sim & L^{\frac{\gamma}{\nu}},
\label{chiL} \\
\xi_L(t_L) & \sim & L,
\label{xiL} \\
G_L(t_L;r) & \sim & r^{-p} \exp{\left({-\frac{r}{N_0L}}\right)} ,
\label{G2L} \\
h_1(t) & \sim & L^{\frac{\Delta}{\nu}}.
\label{YLL} 
\end{eqnarray}
In Eq.(\ref{G2L}), $N_0$ represents an appropriate amplitude and we have omitted the subscript $L$ from the finite-size scaling of the first Lee-Yang zero.

One is also interested in how the location of the peak $t=t_L$ 
is shifted relative to its  infinite-volume limit $t=0$ (the critical point).  
This is characterised by the so-called shift exponent $\lambda_T$.
For a system of linear extent $L$  the scaling of the pseudocritical point also follows a power-law
%
%\mycheck{\red{shift}}
\begin{equation}
 t_L \sim L^{\lambda_T},
 \label{shift}
\end{equation}
to leading order.

The smoothening out of the divergence present in the thermodynamic limit into a peak is also associated with the rounding exponent $\theta_T$.
The rounding may be  defined as the width  of the susceptibility curve at half of its maximum height, $\Delta T$. One then has  
%\mycheck{\red{round}}
\begin{equation}
 \Delta T \sim L^{-\theta_T},
 \label{round}
\end{equation}
to leading order.

%%%%%%%%%%%%%%%%%%%%%%%%%%%%%%%%%%%%%%%%%%%%%%%%%%%%%%%%%%%%%%%%%%%%%%%%%%%
%\mycheck{\red{LeadingR}}
\subsection{Scaling relations}
\label{LeadingR}
%%%%%%%%%%%%%%%%%%%%%%%%%%%%%%%%%%%%%%%%%%%%%%%%%%%%%%%%%%%%%%%%%%%%%%%%%%%
\index{scaling relations}

The six core critical exponents are related by four famous scaling relations,  derived in the 1960's:
%\mycheck{\red{Jo,u,Gr,Fi}}
\begin{eqnarray}
 \nu d & = & 2-\alpha , \label{Jo} \\
 2\beta + \gamma & = & 2 - \alpha ,\label{Ru} \\
 \beta (\delta - 1)  & = & \gamma ,\label{Gr} \\
 \nu (2-\eta)  & = & \gamma . \label{Fi}
\end{eqnarray}
The relation (\ref{Jo}) was developed by Widom\cite{Wi65,Gr67} using  considerations of dimensionality, with alternative arguments given by Kadanoff\cite{Ka66}. 
Josephson\cite{Jo67}\index{scaling relations} later derived the inequality $\nu d \ge 2 - \alpha$ on the basis of some plausible thermodynamic assumptions. 
Because it involves dimensionality, Eq.(\ref{Jo}) is also called the {\emph{hyperscaling relation.}}
It is conspicuous in the set (\ref{Jo})--(\ref{YLe}) in that it is the only scaling relation involving  $d$.
The hyperscaling relation (\ref{Jo}) lies at the heart of this review.
The equality (\ref{Ru}) was originally proposed by Essam and Fisher\cite{EsFi63} and the related inequality $2\beta + \gamma  \ge  2 - \alpha$ was rigorously proven by Rushbrooke\cite{Ru63}.
Similarly,  relation (\ref{Gr}) was put forward  by Widom\cite{Wi64} and the related inequality $\beta (\delta - 1)  \le \gamma$  proved by Griffiths\cite{Gr65}.
Equalities (\ref{Ru}) and (\ref{Gr}) were re-derived by Abe\cite{Ab67p72} and Suzuki\cite{SuzukiLY} using an alternative route involving Lee-Yang zeros.
Eq.(\ref{Fi}) was derived by Fisher\cite{Fi64sc}, with a related inequality proved in Ref.~[\refcite{BuGu69,Fi69inequalities}].
The gap exponent is given by \cite{Domb}
%\mycheck{\red{YLe}}
\begin{equation}
 \Delta  =  \frac{\delta \gamma }{\delta - 1} = \delta \beta = \beta + \gamma \label{YLe}.
\end{equation}
The  critical exponent governing the behaviour of the correlation length in field is
%\mycheck{\red{X}}
\begin{equation}
   \nu_c =  \frac{\nu}{\Delta} . \label{X} 
\end{equation}
The  predictions for the rounding and shifting exponents coming from standard FSS 
%\mycheck{\red{shift2}}
\begin{equation}
 \lambda_T = \frac{1}{\nu} .
\label{shift2}
\end{equation}
But this is not always true and in some cases, such as in the Ising model in two dimensions with special boundary conditions, it can deviate from this value.
Similarly the rounding exponent is
%\mycheck{\red{round2}}
\begin{equation}
\theta_T = \frac{1}{\nu} .
\label{round2}
\end{equation}

%%%%%%%%%%%%%%%%%%%%%%%%%%%%%%%%%%%%%%%%%%%%%%%%%%%%%%%%%%%%%%%%%%%%%%%%%%%
%\mycheck{\red{Logcor}}
\subsection{Logarithmic corrections}
\label{Logcor}
%%%%%%%%%%%%%%%%%%%%%%%%%%%%%%%%%%%%%%%%%%%%%%%%%%%%%%%%%%%%%%%%%%%%%%%%%%%
\index{scaling! logarithmic corrections}
\index{scaling relation! for logarithmic corrections}

In certain circumstances, there are multiplicative logarithmic corrections to the leading behaviour and\cite{KeVol3}
%\mycheck{\red{ctlog,mhlog,chitlog,xitlog}}
\begin{eqnarray}
c_\infty(t)    & \sim & {|t|}^{-\alpha}|\ln{|t|}|^{\hat{\alpha}},            \label{ctlog}\\
m_\infty(t)    & \sim & {(-t)}^{\beta}|\ln{(-t)}|^{\hat{\beta}} \quad {\mbox{for $T<T_c$}},               \label{mtlog}\\
m_\infty(h)    & \sim & {h}^{\frac{1}{\delta}}|\ln{h}|^{\hat{\delta}},   \label{mhlog}\\
\chi_\infty(t) & \sim & {|t|}^{-\gamma}|\ln{|t|}|^{\hat{\gamma}},            \label{chitlog}\\
\xi_\infty(t)    & \sim & {|t|}^{-\nu}|\ln{|t|}|^{\hat{\nu}},                \label{xitlog}\\
r_{\rm{YL}}(t) & \sim & t^{\Delta} |\ln{{t}}|^{\hat{\Delta}}
\quad {\mbox{for $t>0$}}
.
\label{edgelog}
\end{eqnarray}
In addition the scaling of the correlation function\index{correlation function} at $h=0$ has
%
%\mycheck{\red{corrfunlog}}
\begin{equation}
 {\cal{G}}_\infty (r,t) \sim r^{-(d-2+\eta)}(\ln{r})^{\hat{\eta}}
 D\left[
 \frac{r}{\xi_\infty(t)}
 \right]
 ,
 \label{corrfunlog}
\end{equation}
To allow for logarithmic corrections in the correlation length of the finite-size system, we write
%
%\mycheck{\red{corrLlog}}
\begin{equation}
  \xi_L(0) \sim L (\ln{L})^{\hat{\qq}}
  .
  \label{corrLlog}
\end{equation}
Recently scaling relations between these logarithmic-correction exponents have been established \cite{KJJ2006} 
Analogues of Eqs.(\ref{Jo})--(\ref{Fi}) are (for a review see the previous volume in this series \cite{KeVol3})
%
%\mycheck{\red{SRlog1,SRlog2,SRlog3,SRlog4}}
\begin{eqnarray}
  \hat{\alpha} & = & \left\{{\begin{array}{l}
                                             ~ 1 + d (\hat{\qq} -  \hat{\nu}) \quad  {\mbox{if}} \quad \alpha = 0 \quad {\mbox{and}} \quad \phi \ne \pi/4 \\
                                             ~ d (\hat{\qq} -  \hat{\nu})   \quad  {\mbox{otherwise,}}
                     \end{array}}\right.
  \label{SRlog1}  \\
   2 \hat{\beta} - \hat{\gamma}  & = &  d(\hat{\qq}-{\hat{\nu}}) ,
  \label{SRlog2}  \\
  \hat{\beta} (\delta - 1) & = &   \delta \hat{\delta} - \hat{\gamma} ,
  \label{SRlog3} \\
  \hat{\eta}  & = &  \hat{\gamma} - \hat{\nu} (2 - \eta ) .
  \label{SRlog4} 
\end{eqnarray}
In the first of these, $\phi$ refers to the angle at which the Fisher zeros impact onto the real axis.
If $\alpha = 0$, and if this impact angle\index{impact angle} is any value other than $\pi/4$, an extra logarithm arises in the specific heat.
For example, this happens in $d=2$ dimensions, but not in $d=4$, where $\phi = \pi/4$ \cite{KJJ2006}.
%Widom\cite{Wi65} also showed how  a logarithmic singularity can arise in the specific heat if $\alpha=0$, but does not have to, leaving instead a finite  discontinuity (see also Ref.~[\refcite{Fi65,Gr67}]).
One notes the crucial role played by the exponent $\hat{\q}$ in the scaling relations for logarithmic corrections.
The question  arises, what is the analogue of this for the leading exponents, and why does it not appear in the usual scaling relations (\ref{Jo})--(\ref{Fi}).
That is the subject of much of what follows and next we summarise the answer.  

%%%%%%%%%%%%%%%%%%%%%%%%%%%%%%%%%%%%%%%%%%%%%%%%%%%%%%%%%%%%%%%%%%%%%%%%%%%
\subsection{$Q$-Scaling and $Q$-FSS: The new paradigm}
%\mycheck{\red{Qnew}}
\label{Qnew}
%%%%%%%%%%%%%%%%%%%%%%%%%%%%%%%%%%%%%%%%%%%%%%%%%%%%%%%%%%%%%%%%%%%%%%%%%%%
\index{scaling! $Q$-scaling}

For clarity and convenience, we gather here the new results recently derived for scaling and FSS above the upper critical dimension and reviewed in this Chapter.

Of core importance is the leading power-law analogue of $\hat{\q}$ in the correlation length, 
\begin{equation}
 \xi_L (t_L) \sim L^{\qq}.
\end{equation}
We shall show that the new exponent $\q$ is given by
\begin{equation} 
  \q \sim 
\left\{ \begin{array}{ll}
                                \frac{d}{d_c} & \quad \mbox{if $d > d_c$ } \\
                                           1  & \quad \mbox{if $d< d_c$} \\
                                                  \end{array}
                                                       \right..
%\label{resut1}
\end{equation}
We will also establish the leading power-law analogue to Eq.(\ref{SRlog1}) as 
\begin{equation}
 \frac{\nu d}{\q} = 2- \alpha.
\end{equation} 
This scaling relation holds in all dimensions and is the extension of hyperscaling to $d>d_c$.

We will also show that  two separate correlation functions are required above $d_c$. 
Which one to use depends upon whether one uses the scale of the lattice extent $L$ or  the correlation length $\xi_L$. In the former case  one has at criticality
\begin{equation}
 G_L (r) \sim r^{-d-2+\eta_Q},
\end{equation}
while in the latter case,
\begin{equation}
 G_\xi (r) \sim r^{-d_c-2+\eta}.
\end{equation}
The formula (\ref{G1}) is therefore only valid below $d_c$.
The second new exponent $\eta_Q$ is  
the anomalous dimension when distance is measured in the scale of $L$.
It is related to $\eta$, the anomalous dimension on the scale of $\xi$, by
\begin{equation}
 \eta_Q = \q \eta + 2(1-\eta).
\end{equation}
Thus $\eta_Q$ and $\eta$ coincide when $d<d_c$.
These anomalous dimensions also have logarithmic counterparts for the case when $d=d_c$, thus providing an additional formula to the list (\ref{SRlog1})--(\ref{SRlog4}) and an amendment to Chapter~1 of Ref.~[\refcite{KeVol3}].

In the following, we provide evidence that the new  exponents  are both {\emph{physical}} and {\emph{universal}}. 
They are physical in the sense that they control the finite-size behaviour of the correlation length
and correlation function.
We also suggest that they are universal, independent of the boundary conditions used for finite-size systems .

%%%%%%%%%%%%%%%%%%%%%%%%%%%%%%%%%%%%%%%%%%%%%%%%%%%%%%%%%%%%%%%%%%%
%\mycheck{\red{11.05.2009c}}
\section{Widom Scaling and Kadanoff Renormalization as Bases for the Scaling Relations}
\label{11.05.2009c}
\setcounter{equation}{0}
%%%%%%%%%%%%%%%%%%%%%%%%%%%%%%%%%%%%%%%%%%%%%%%%%%%%%%%%%%%%%%%%%%%

\index{Widom scaling}

Historically, scaling theory begins with Widom's scaling ansatz for the magnetisation \cite{Wi65}
(see also Ref.~[\refcite{PP}])
%\mycheck{\red{CM2.165}}
\begin{equation}
 m_\infty(t,h) = |t|^{\beta}{\mathcal{M}}_{\pm}\left({\frac{h}{|t|^{\Delta}}}\right), \quad
\mbox{for} \quad t \rightarrow 0^{\pm}, h \rightarrow 0.
\label{CM2.165}
\end{equation}
Setting $|t| = h^{1/\Delta}$ allows one to re-express this as $ m_\infty(t,h) \sim h^{\beta/\Delta}$. Comparing with Eq.(\ref{mt}), one identifies the gap exponent as 
%\mycheck{\red{gap}}
\begin{equation}
 \Delta = \beta \delta .
\label{gap}
\end{equation}
To achieve the form (\ref{CM2.165}), Widom suggested that the singular part of the free energy scale as
%\mycheck{\red{CM2.168}}
\begin{equation}
 f_\infty(t,h) = |t|^{2-\alpha} {\mathcal{F}}_{\pm} \left({\frac{h}{|t|^{\Delta}}}\right).
\label{CM2.168}
\end{equation}
(This functional form of the scaling function ensures that $h$ enters the partition function through the ratio 
$h/|t|^{\Delta}$. This means that the Lee-Yang zeros scale as $|t|^{\Delta}$, so that $\Delta$ is the gap exponent of Eq.(\ref{YL}).)
A similar scaling ansatz can be written for the correlation function \cite{Ka66}:
%\mycheck{\red{CM2.182}}
\begin{equation}
 G_\infty(r,t,h) \sim \frac{1}{r^{d-2+\eta}} {\mathcal{G}}_{\pm}\left(\frac{r}{\xi},{\frac{h}{|t|^{\Delta}}}\right), \quad
\mbox{for} \quad t \rightarrow 0^{\pm}, h \rightarrow 0.
\label{CM2.182}
\end{equation}
Together with the assumption that the free energy scale as the inverse correlation volume 
%\mycheck{\red{CM2.189}}
\begin{equation}
 f_\infty(t,h) \sim \xi_\infty^{-d},
\label{CM2.189}
\end{equation}
one has a complete description of scaling in the thermodynamic limit, from which the scaling relations (\ref{Jo})--(\ref{Fi})  follow.

Firstly, Eq.(\ref{CM2.168}) with $h$ set to zero and Eq.(\ref{CM2.189}) deliver the hyperscaling relation $\nu d = 2 - \alpha$.
Next, differentiating Eq.(\ref{CM2.168}) with respect to field gives $m_\infty (t) \sim |t|^{2-\alpha - \Delta} $, from which we have $\Delta = \beta + \gamma$. Combined with Eq.(\ref{gap}) this gives Widom's relation (\ref{Gr}).
Differentiating a second time gives $\chi_\infty (t) \sim |t|^{2-\alpha - 2\Delta} $.
Identifying the exponent as $-\gamma$ delivers the relation (\ref{Ru}).

The starting point for the standard derivation of  Fisher's relation (\ref{Fi})
is the fluctuation-dissipation theorem
%\mycheck{Dissresp}
\begin{equation} 
 \chi_\infty (t)  =  \int_0^{\xi_\infty(t)}{d^d\mathbf{r}\ \! G_\infty(r,t,0)},
\label{Dissresp}
\end{equation}
having bounded the integral by the correlation length.
Then, from the form~(\ref{CM2.182}),
%\mycheck{Fisherreln1}
\begin{equation}
 \chi_\infty (t)= \int_0^{\xi_\infty(t)}{dr \ \!r^{1-\eta}\ \! }
								       \mathcal{G}_{\pm} \left[{\frac{r}{\xi_\infty(t)}}\right].
\label{Fisherreln1}											
\end{equation}
Fixing $r/\xi_\infty(t) = x$, this gives
%\mycheck{Fisherreln2}
\begin{equation}
 \chi_\infty (t)= \xi_\infty^{2-\eta} 
                  \int_0^1{dx\ \! x^{1-\eta} }
								       \mathcal{G}_{\pm} \left({x}\right).
\label{Fisherreln2}		
\end{equation}
Finally, inserting the scaling behaviour for $\xi_\infty$ and $\chi_\infty$, one obtains Fisher's relation\cite{Fi64sc} $\gamma = \nu (2-\eta)$.

We will revisit this derivation in Sec.\ref{EPLF} where we will see that this 50-year old scaling relation needs to be modified above the upper critical dimension.

\index{Kadanoff block spins}

The Widom scaling ansatz may be justified through  Kadanoff's block spin renormalization approach.\cite{Ka66}
One partitions the lattice into blocks of size $b$ (in units of $a$) and replaces each of the $b^d$ spins in a block by a single block spin $s_I$. 
One then rescales all lengths  by an amount $b$.
At the critical point $(t,h)=(0,0)$  (recall $t=T/T_c-1$ is the reduced temperature), the correlation length is infinite and remains so after the block spin transformation -- its is a fixed point of the transformation which maps $(t,h)$ to new values $(t^\prime,h^\prime)$.
Near the critical point, the relationship between the original and renormalized parameters is $t^\prime = \lambda_t(b)t$ and $h^\prime = \lambda_h(b)h$.
Demanding that successive blocking is equivalent to a single transformation, $\lambda_i(b_2) \lambda_i(b_1) = \lambda_i (b_1b_2)$ and that the identity transformation effect no change, delivers the expectation that $t^\prime = b^{y_t}t$ and $h^{\prime} = b^{y_h} h$ with $y_i > 0$ for $i =t,h$.
The first of these then gives  $
 \xi_\infty^\prime(t) \sim {|t^\prime|}^{-\nu} 
\sim b^{-y_t \nu} \xi_\infty(t)$.
But since in Kadanoff's approach the correlation length is transformed as $\xi_\infty^\prime = \xi_\infty/b$, we identify
%\mycheck{\red{CM2.221}}
\begin{equation}
 \nu = \frac{1}{y_t}.
\label{CM2.221}
\end{equation}

Demanding that the partition function remains unchanged under the real-space renormalization transformation $Z_L(t,h) = Z_{L^\prime}(t^\prime,h^\prime)$, where $L^\prime  =L/b$
(here, $N=L^d$ is the number of original spins and $N^\prime= {L^{\prime}}^d$ is the number of block spins), delivers for the free energy in the critical region,
%\mycheck{\red{CM2.227}}
\begin{equation}
 f_{\infty}(t,h) = b^{-d} f_{\infty}(b^{y_t}t,b^{y_h}h).
\label{CM2.227}
\end{equation}
This is a generalised homogeneous function.
From this, Widom's scaling follows as 
\[
 f_{\infty}(t,h) = |t|^{\nu d}f_{\infty}(\pm 1,h|t|^{-y_h/y_t}).
\]
Comparing to Eq.(\ref{CM2.168}), one has $\nu d = 2 - \alpha$ and 
\begin{equation}
 \Delta = \frac{y_h}{y_t}
 \quad 
 {\mbox{or}}
 \quad
 y_h = \frac{\beta \delta}{\nu}.
\end{equation}
Therefore the assumption that the renormalized partition function take the same form as the original one delivers a generalised homogeneous free energy, which then delivers Widom's ansatz and  hyperscaling.

Kadanoff's block spin technique can also be applied to the correlation function.
Write
${{
 m_I = b^{-d} \sum_{i \in I}^{b^d}{s_i}
}}$
and define the block spin $
 S_I = {\rm{sign}}(m_I) = {m_I}/{|m_I|}$..
Then,
\begin{eqnarray*}
 G_{\infty}(r^\prime, t^\prime, h^\prime) & = & 
 \left\langle{S_IS_J}\right\rangle- \left\langle{S_I}\right\rangle \left\langle{S_J}\right\rangle \\
 & = & \frac{1}{|m_I||m_J|} 
       \frac{1}{b^{2d}}
			 \sum_{i \in I}^{b^d}\sum_{j \in J}^{b^d}{
			  \left\langle{s_is_j}\right\rangle- \left\langle{s_i}\right\rangle \left\langle{s_j}\right\rangle } \\
				& = & 
				\frac{1}{|m_I||m_J|} G_{\infty}(r,t,h).
\end{eqnarray*}
From FSS, $m_I \sim b^{-\beta/\nu}$, so we have
 $G_{\infty}(r^\prime, t^\prime, h^\prime) \sim b^{\beta/\nu}G_{\infty}(r,t,h)$ or
%\mycheck{\red{CM2.236}}
\begin{equation}
 G_{\infty}(r,t,h) \sim b^{-\frac{\beta}{\nu}} G_{\infty}(rb^{-1}, tb^{y_t}, hb^{y_h}) .
\label{CM2.236}
\end{equation}
The scaling ansatz (\ref{CM2.182}) follows from this.

%%%%%%%%%%%%%%%%%%%%%%%%%%%%%%%%%%%%%%%%%%%%%%%%%%%%%%%%%%%%%%%%%%%%%%%%%%%
%\mycheck{\red{III}}
\section{Mean-Field Theory, Landau Theory, $\phi^4$ Theory and the Ginzburg Criterion}
\label{III}
\setcounter{equation}{0}
%%%%%%%%%%%%%%%%%%%%%%%%%%%%%%%%%%%%%%%%%%%%%%%%%%%%%%%%%%%%%%%%%%%%%%%%%%%

The mean-field theory\cite{Curie1895,Weiss1907}  for the Ising model and Landau theory\cite{Landau37} are both historically and conceptually the basis for deeper, firmer and more realistic theories of critical phenomena.
They also produce critical exponents which obey the scaling relations (except hyperscaling).
Here we present both theories along with a brief account of a criterion which marks their validity.

%%%%%%%%%%%%%%%%%%%%%%%%%%%%%%%%%%%%%%%%%%%%%%%%%%%%%%%%%%%%%%%%%%%%%%%%%%%
\subsection{Mean-field theory for the Ising model}
%%%%%%%%%%%%%%%%%%%%%%%%%%%%%%%%%%%%%%%%%%%%%%%%%%%%%%%%%%%%%%%%%%%%%%%%%%%

\index{mean-field theory}

The Ising Hamiltonian is given by Eqs.(\ref{II2}), (\ref{II10}) and (\ref{II11}),
%\mycheck{\red{LBpage12}}
\begin{equation}
 \mathcal{H} = - J \sum_{\langle{i,j}\rangle}{s_is_j} -  H \sum_i{s_i}.
 \label{LBpage12}
\end{equation}
Writing $s_i = m_L + \delta s_i$ in the first term, in which $m_L = \langle{s_i}\rangle$, 
\begin{equation}
 \mathcal{H} = - J \sum_{\langle{i,j}\rangle}{(m_L+\delta s_i)(m_L+\delta s_j)} -  H \sum_i{s_i}.
\end{equation}
(In writing $m_L$ as independent of $i$, we have again assumed translational invariance for simplicity.)
We express this as
%\mycheck{\red{CM2.206}}
\begin{equation}
 \mathcal{H} = \mathcal{H}_{\rm{MF}} + \Delta \mathcal{H} , 
\label{CM2.206}
\end{equation}
where
\begin{equation}
 \mathcal{H}_{\rm{MF}}
  = 
  Jm^2 \sum_{\langle{i,j}\rangle}{1} - 2Jm_L \sum_{\langle{i,j}\rangle}{s_i} - H \sum_i{s_i},
\end{equation}
having reinstating $s_i$, and
%\mycheck{\red{DeltaH}}
\begin{equation}
 \Delta \mathcal{H} = - J\sum_{\langle{i,j}\rangle}{(\delta s_i)(\delta s_j)}.
\label{DeltaH}
\end{equation}
The mean-field approximation consists of neglecting second-order fluctuations so that the energy of the model is simply $\mathcal{H}_{\rm{MF}}$. 
%The validity of this approximation is addressed in Sec.~\ref{secGinzburg}.

Introduce the coordination number $q$ as the number of nearest neighbours.
For example, a hypercubic lattice with periodic boundary conditions has $q=2d$. 
Then $\sum_{\langle{i,j}\rangle}{1} = qN/2$
and  $\sum_{\langle{i,j}\rangle}{s_i}  = (q/2) \sum_i{s_i}$, so that
%\mycheck{\red{III3i}}
\begin{equation}
 \mathcal{H}_{\rm{MF}} 
 %& = & - \frac{J}{2} \sum_i{\sum_{j=1}^q}{(2ms_i-m^2)} -  m \sum_i{s_i},  \label{III2} \\
  =  \frac{q}{2}NJm_L^2 - (Jqm_L+H) \sum_i{s_i}.
 \label{III3i}
\end{equation}
The Hamiltonian has  been decoupled into a sum of single-body, non-interacting effective Hamiltonians in an effective mean-field  $H_{\rm{eff}}= H + Jqm_L$. 
Now sum over the configurations and insert into the partition function to obtain
%\mycheck{\red{III5}}
\begin{equation}
 Z_{\rm{MF}}  =  \sum_{\{s_i\}}{e^{-\beta{\mathcal{H}}_{\rm{MF}}}}
  =  e^{-\frac{\beta N q J m_L^2}{2}  } [2\cosh{ (\beta H + J\beta qm_L)}]^N.
 \label{III5}
\end{equation}
From Eq.(\ref{II3}), the free energy is then
%\mycheck{\red{CM2.101}}
\begin{equation}
 f_{\rm{MF}} = \frac{qJm_L^2}{2} - k_BT\ln{[2\cosh{ (\beta H + J\beta qm_L)}]}.
\label{CM2.101}
\end{equation}
%Note that $m$ is not an independent variable here - it is determined by minimising $f$.
Eq.(\ref{II8}) then gives the transcendental equation 
%Le Bellac page 13
%\mycheck{\red{III6}}
\begin{equation}
 m_L = \tanh{[\beta (H + Jqm_L)]},
 \label{III6}
\end{equation}
or
%\mycheck{\red{III7}}
\begin{equation}
 \tanh^{-1}{m_L} = \frac{1}{2} \ln{\left({\frac{1+m_L}{1-m_L}}\right)} = \beta H + \beta Jqm_L .
  \label{III7} 
\end{equation}
Expanding the inverse hyperbolic function,
%\mycheck{\red{III7a}}
\begin{equation}
  m_L + \frac{m_L^3}{3} + \dots = \beta H + \beta Jqm .
  \label{III7a} 
\end{equation}
When $h=\beta H=0$, this delivers the solutions $m_L=0$, which can hold for any $T$, 
and $m_L = \pm \sqrt{3(\beta Jq-1)}$, which is real only for $\beta \ge 1/Jq$.
Identify
%\mycheck{\red{MFTc}}
\begin{equation}
 \beta_c = \frac{1}{qJ} \quad {\mbox{or}} \quad T_c = \frac{qJ}{k_B} .
\label{MFTc}
\end{equation}
There is therefore a phase transition from a zero-magnetisation phase when  $H=0$ and $T>T_c$ to a magnetised phase when $H=0$ and $T<T_c$. 
These solutions coincide at $T_c$.

Differentiating Eq.(\ref{III7a}) with respect to $h$ gives $\chi_L (1-\beta / \beta_c) + m_L^2 \chi_L + \dots = \beta $, leading to $\gamma = 1$ in the thermodynamic limit.
The critical isotherm $\beta = \beta_c$ gives $m_L^3 = 3 \beta_c H$ so that $\delta = 3$.
Finally, the internal energy is the expectation value of the Hamiltonian in Eq.(\ref{II15}).
Differentiating with respect to temperature then delivers $\alpha =0$.

The  mean-field theory  delivers phase transitions even for finite $L$ and even in in one dimension.
However, there can be no genuine transition in a finite-size system and, 
according to Ising's calculation,  there should be no spontaneous magnetisation in $d=1$ \cite{Ising}.  
At the other extreme, mean-field theory becomes exact in the limit where the
interactions are between all pairs of spins and not just nearest neighbours.
It is also exact in the limit of infinite dimensionality.
In summary, mean-field theory leads to the prediction 
$\alpha = 0$, $\beta = 1/2$, $\gamma = 1$ and $\delta = 3$, independent of dimensionality.

Rather than expanding out Eq.(\ref{III6}), one can expand the more fundamental equation (\ref{CM2.101}).
One finds (dropping the subscripts $L$),
%\mycheck{\red{CM2.128}}
\begin{equation}
f_{\rm{MF}} = f_0 - Hm + a_2(T-T_c)m^2 + a_4 m^4 + \dots,
\label{CM2.128}
\end{equation}
where $f_0 = - k_BT \ln 2$, $a_2 = k_B/2$ and $a_4 = k_BT/12$.
When $T>T_c$, the order parameter vanishes, so that $f_{\rm{MF}} = f_0$.

One can now proceed to determine $m$, $\chi$ and $c$ in the usual manner and verify that the Taylor expansion of the mean field delivers the same scaling behaviour in the vicinity of the critical point as the full mean-field theory.

%%%%%%%%%%%%%%%%%%%%%%%%%%%%%%%%%%%%%%%%%%%%%%%%%%%%%%%%%%%%%%%%%%%%%%%%%%%
%\mycheck{\red{IIIsub2}}
\subsection{Landau theory}
\label{IIIsub2}
%%%%%%%%%%%%%%%%%%%%%%%%%%%%%%%%%%%%%%%%%%%%%%%%%%%%%%%%%%%%%%%%%%%%%%%%%%%

\index{Landau theory}

Eq.(\ref{CM2.128}) coincides with Landau's phenomonological approach to phase transitions.
That approach  is to identify the order parameter $\phi_0$ and its symmetries and then to construct a Hamiltonian from all possible invariants subject to spatial or space-time symmetries\cite{Landau37}.
The Ising model has symmetry under $\phi_0 \rightarrow -\phi_0$, so that the polynomial should contain only even powers. 
The idea behind Landau's approach is that if the order parameter is small near the phase transition, the free energy can be expanded in powers of $\phi_0$.
That expansion can then be truncated close enough to the transition point. 
The coefficients of the power series are functions of the control parameters $T$ and $H$.
For the symmetries of the Ising model then, the first few terms are
given by $
 \beta f(t,h;\phi_0) 
 = \beta f_0(t,h)+[{r_0(t,h)}/{2}] \phi_0^2(t,h) + [{u(t,h)}/{4}] \phi_0^4(t,h) - h \phi_0(t,h)
$.
Here we have expanded $\beta f$ instead of $f$ for later convenience when we connect with $\phi^4$ theory. %I do this
We have also introduced the expansion coefficients as fractions for  the same reason.
We next absorb $\beta \approx \beta_c$ into $f$ and write this as 
%\mycheck{\red{rain1}}
\begin{equation}
  f(t,h;\phi_0) 
 =  f_0(t,h)+\frac{r_0(t,h)}{2} \phi_0^2(t,h) + \frac{u(t,h)}{4} \phi_0^4(t,h) - h \phi_0(t,h) .
\label{rain1}
\end{equation}
Minimising Eq.(\ref{rain1}) with respect to the order parameter $\phi_0$, one  obtains
%\mycheck{\red{min1,min2}}
\begin{eqnarray}
 \frac{\delta f}{\delta \phi_0} & = & r_0(t,h)\phi_0 + u(t,h) \phi_0^3 - h =0
,
\label{min1}
\\
 \frac{\delta ^2 f}{\delta \phi_0^2} & = & r_0(t,h) + 3 u(t,h) \phi_0^2 > 0
.
\label{min2}
\end{eqnarray}

If $h=0$, and if $r_0(T)>0$, Eq.(\ref{min1}) can only hold if $\phi_0=0$, so we identify this as the symmetric phase ($T>T_c$).
If $r_0<0$, the  equation permits the solution
\begin{equation}
 \phi_0 = \sqrt{-\frac{r_0(T)}{u(T)}}
,
\label{MFM}
\end{equation}
which we can associate with the broken ($T<T_c$) phase. 
In both cases Eq.(\ref{min2}) is satisfied.
Since $r_0(T)$ changes sign at $T_c$, its expansion in terms of $T$ should take the form
%\mycheck{\ref{ritot}}
\begin{equation}
 r_0(T) = {r_0}_1t + {r_0}_3t^3 + \dots
,
\label{ritot}
\end{equation}
where ${r_0}_1>0$. Similarly expanding 
\begin{equation}
u(T) = u_0 + u_1t + \dots
,
\end{equation}
with $u_0>0$, Eq.(\ref{MFM}) becomes
\begin{equation}
 \phi_0 = \sqrt{\frac{{r_0}_1}{u_0} } |t|^{\frac{1}{2}} + \dots, {\mbox{for $t<0$.}}
\label{MFM2}
\end{equation}
Since $\phi_0  $ is the order parameter in the Landau theory, we can identify the mean-field critical exponent $\beta = 1/2$.

The specific heat is $c = -T \partial^2f / \partial T^2 \sim - \partial^2f/\partial t^2 $.
Firstly,
\[
 \frac{\partial f}{\partial t} 
 =
 \frac{r_0^\prime}{2}\phi_0^2 + r_0 \phi_0 \phi_0^\prime + \frac{u^\prime}{4}\phi_0^4 + u \phi_0^3 \phi_0^\prime,
\]
with primes indicating derivatives with respect to $t$.
Now, the leading terms in the expansions of $u$ and $u^\prime$ are constant, while $\phi_0^\prime$ brings in a term proportional to $|t|^{-1}$, so we  derivatives of $u$ give sub-leading terms and can be dropped. (We can treat $u$ as a $t$-independent parameter.) This gives
\[
 -c = \frac{\partial^2 f}{\partial t^2} 
 =
 2r_{01} \phi_0 \phi_0^\prime + r_{01} (\phi_o^\prime)^2 + r_{01} \phi_0 \phi_0^{\prime \prime} + 3 u_0 \phi_0^2 (\phi_0^\prime)^2 + u_0 \phi_0^3 \phi_0^{\prime \prime}.
\]
We use $\phi_0=0$ above $T_c$ and Eq.(\ref{MFM}) below $T_c$ to arrive at
\begin{equation} 
  c \sim 
\left\{ \begin{array}{ll}
                                                                0 & \mbox{if $t > 0$ } \\
                                                                   ~ & ~ \\
                                                                 \frac{r_0^2}{2u_0}  & \mbox{if $t < 0$} \\
                                                  \end{array}
                                                       \right.
\label{cMF}
\end{equation}
Therefore we identify  $\alpha= 0$.

If $t=0$ or $r_0=0$ in Eq.(\ref{min1}), we have 
\begin{equation}
 \phi_0^3 = \frac{h}{u}
,
\label{deltais3}
\end{equation}
 yielding $\delta = 3$.

Differentiating (\ref{min1}) with respect to $h$ and identifying the susceptibility as $\chi = \partial \phi_0 / \partial h$
gives, in the absence of the external field,
\begin{equation} 
  \chi 
 =
\left\{ \begin{array}{ll}
                                                                 \frac{1}{{r_0}(t)}  & \mbox{if $t > 0$ } \\
                                                                   ~ & ~ \\
                                                                 -\frac{1}{{2r_0}(t)}  & \mbox{if $t < 0$} \\
                                                  \end{array}
                                                       \right. .
\label{ggg}
\end{equation}
Thus we identify $\gamma=1$.

To obtain the correlation function, we require the Ornstein-Zernike extension \cite{OZ}
\index{Ornstein-Zernike}
of the Landau theory (\ref{rain1})  
%\mycheck{\red{IrlF20}}
\begin{equation}
 F[\phi_0,h] 
 = {\int{d^d{\mathbf{r}}\ \!{\left[{\frac{1}{2}[\nabla \phi_0 ({\mathbf{r}})]^2 + \frac{r_0}{2} \phi_0^2({\mathbf{r}}) + \frac{u}{4} \phi_0^4({\mathbf{r}}) - h({\mathbf{r}}) \phi_0({\mathbf{r}})}\right]}}}.
\label{IrlF20}
\end{equation}
(A general coefficient of the $(\nabla \phi_0)^2$ term here can be incorporated into the remaining coefficients. Here we set it to $1/2$ for later convenience.)
Minimising,
%\mycheck{\red{hoover1}}
\begin{equation}
 \frac{\delta F}{\delta \phi_0({\mathbf{r}})}
 =
 r_0 \phi_0({\mathbf{r}}) + u \phi_0^3({\mathbf{r}}) - h ({\mathbf{r}}) - \nabla^2 \phi_0({\mathbf{r}}) = 0.
\label{hoover1}
\end{equation}
Differentiate the associated Eq.(\ref{hoover1}) with respect to field $h({\mathbf{r}}^\prime)$ at location ${\mathbf{r}}^\prime$ to obtain
\begin{equation}
 r_0 G({\mathbf{r}},{\mathbf{r}}^\prime) + 3 u \phi^2({\mathbf{r}}) G({\mathbf{r}},{\mathbf{r}}^\prime)
  - \nabla^2 G({\mathbf{r}},{\mathbf{r}}^\prime) = \delta ({\mathbf{r}} - {\mathbf{r}}^\prime).
\label{MFT11}
\end{equation}
If the field $h$ is uniform, translational invariance means that 
$G({\mathbf{r}},{\mathbf{r}}^\prime) = G(|{\mathbf{r}}-{\mathbf{r}}^\prime|)$.
We find
%\mycheck{\red{}{MFT1211}}
\begin{equation}
 - \nabla^2 G(r) +  R_0 G(r)  =  \delta(r) ,
\label{MFT1211}
\end{equation}
where $R_0 = r_0$ if $T>T_c$,  $R_0 = - 2 r_0$ if $T<T_c$ and $R_0 = 0$ if $T=T_c$.

The Fourier transform of Eq.(\ref{MFT1211}) is 
%\mycheck{\red{LB2.3.3'}}
\begin{equation}
  (k^2 + R_0)\tilde{G}(k) = 1.
\label{LB2.3.3'}
\end{equation}
The solution of Eq.(\ref{LB2.3.3'}) is the Ornstein-Zernike form,\cite{OZ}
%\mycheck{\red{LB2.3.3}}
\begin{equation}
 \tilde{G}(k) = \frac{1}{k^2 +  \xi^{-2}},
\label{LB2.3.3}
\end{equation}
which is exact for Landau theory. Here 
\begin{equation}
 \xi = R_0^{-\frac{1}{2}}
\end{equation}
 is the correlation length from Eq.(\ref{G1mon}).
The inverse Fourier transform is then
%\mycheck{strct}
\begin{equation}
 G(r) \sim \frac{1}{r^{d-2}} g\left({\frac{r}{\xi}}\right),
\label{strct}
\end{equation}
where 
\begin{equation}
 g\left({\frac{r}{\xi}}\right) =\left({\frac{r}{\xi}}\right)^{\frac{d}{2}-1}
 K_{\frac{d}{2}-1}\left({\frac{r}{\xi}}\right)
\end{equation}
in which $K_{\frac{d}{2}-1}$ is a modified Bessel function.
Therefore
%\mycheck{\red{sunn0}}
\begin{equation}
 G(r) \sim \frac{\xi^{1-d/2}}{r^{d/2-1}} K_{\frac{d}{2}-1}\left({\frac{r}{\xi}}\right).
\label{sunn0}
\end{equation}
Now, $K_{\nu}(x) \sim 1/x^{\nu} $ as $x \rightarrow 0$, so when $\xi \rightarrow \infty$, 
\begin{equation}
 G(r) \sim \frac{1}{r^{d-2}} \quad {\mbox{or}} \quad \eta=0,
\end{equation}
the correlation length having dropped out.
For finite $\xi \ll r$, the asymptotic behaviour $
 K_\nu(x) \rightarrow e^{-x}/\sqrt{x}$ for large $x$
gives
\begin{equation}
 G(r) \sim \frac{1}{r^{(d-1)/2}}e^{-r/\xi}.
 \label{long}
\end{equation}

%%%%%%%%%%%%%%%%%%%%%%%%%%%%%%%%%%%%%%%%%%%%%%%%%%%%%%%%%%%%%%%%%%%%%%%%%%%
%\mycheck{\red{LGWtheory}}
\subsection{The Ginzburg-Landau-Wilson $\phi^4$ Theory}
\label{LGWtheory}
%%%%%%%%%%%%%%%%%%%%%%%%%%%%%%%%%%%%%%%%%%%%%%%%%%%%%%%%%%%%%%%%%%%%%%%%%%%

\index{$\phi^4$ theory}

To go beyond mean-field theory or Landau theory, we need to be able to take into account fluctuations in the field $\phi({\mathbf{r}})$. 
The connection between the Ising model and  $\phi^4$ theory is  established through the renormalization group. The Ginzburg-Landau-Wilson 
 partition function is
%\mycheck{\red{Amit6.7}}
\begin{equation}
 Z[h] = \int{{\mathcal{D}}\phi \exp{(-S[\phi])}},
\label{Amit6.7}
\end{equation}
 where the action is
%\mycheck{\red{Amit6.7b}}
\begin{equation}
 S[\phi] = {\int{d^d{\mathbf{r}}\ \,{\left\{{\frac{1}{2}[\nabla \phi ({\mathbf{r}})]^2 + \frac{r_0}{2} \phi^2({\mathbf{r}}) + \frac{u}{4} \phi^4({\mathbf{r}}) - h({\mathbf{r}}) \phi({\mathbf{r}})}\right\}}}},
\label{Amit6.7b}
\end{equation}
having also included a source or field $h({\mathbf{r}})$. 
The functional integration in Eq.(\ref{Amit6.7}) is over continuously fluctuating fields $\phi$.
We write the free energy as
%\mycheck{\red{LBpage84}}
\begin{equation}
 F[h] = - \ln Z[h].
\label{LBpage84}
\end{equation} 
%The susceptibility is given by the second derivative of of the free energy with respect to field $h$ (see Eq.(\ref{II9})) and its positivity shows that the $F$ is concave.
Since $F[h]$ is concave, we define the convex functional
%\mycheck{\red{LB2.1.6}}
\begin{equation}
 W[h] = - F[h] = \ln{Z[h]}.
\label{LB2.1.6}
\end{equation}
From Eq.(\ref{II8}), the magnetisation  is 
%\mycheck{\red{cyr}}
\begin{equation}
 m({\mathbf{r}}) = \frac{\delta W[h]}{\delta h({\mathbf{r}}) } = \langle{\phi({\mathbf{r}})}\rangle .
\label{cyr}
\end{equation}
Also, from Eq.(\ref{II14}), the connected correlation function is
%\mycheck{\red{Gme}}
\begin{equation}
 G({\mathbf{r}}-{\mathbf{r}}^\prime) = \frac{\delta^2 W}{\delta h({\mathbf{r}})\delta h({\mathbf{r}}^\prime)} = \frac{\delta m({\mathbf{r}})}{\delta h({\mathbf{r}}^\prime)}.
\label{Gme}
\end{equation}

The associated Legendre transformation 
%\mycheck{\red{LBpage26}}
\begin{equation}
 \Gamma[m] = \int{d^d{\mathbf{r}}\ \, [m({\mathbf{r}}) h({\mathbf{r}})]} - W[h]
\label{LBpage26}
\end{equation}
is called the Gibbs free energy.  \index{energy!Gibbs free}  
Its usefulness is linked to the correlation function. 
Differentiating Eq.(\ref{Gme}),
\[
 \frac{\delta \Gamma[m]}{\delta m({\mathbf{r}})} = h({\mathbf{r}}) + m \frac{\delta h}{\delta m} - \frac{\delta W}{\delta h} \frac{\delta h}{\delta m},
\]
the last two terms cancel due to Eq.(\ref{cyr}).
Therefore
%\mycheck{\red{LBpage26a}}
\begin{equation}
 \frac{\delta \Gamma[m]}{\delta m({\mathbf{r}})} = h({\mathbf{r}}) .
\label{LBpage26a}
\end{equation}
Differentiating again,
%\mycheck{\red{LBpage26b}}
\begin{equation}
 \frac{\delta^2 \Gamma[m]}{\delta m({\mathbf{r}})\delta m({\mathbf{r}}^\prime)} =  \frac{\delta h({\mathbf{r}})}{\delta m({\mathbf{r}}^\prime)} = G^{-1}({\mathbf{r}},{\mathbf{r}}^\prime),
\label{LBpage26b}
\end{equation}
from Eq.(\ref{Gme}).

%%%%%%%%%%%%%%%%%%%%%%%%%%%%%%%%%%%%%%%%%%%%%%%%%%%%%%%%%%%%%%%%%%%
\subsection{Ginzburg criterion}
\label{secGinzburg}
%\setcounter{equation}{0}
%%%%%%%%%%%%%%%%%%%%%%%%%%%%%%%%%%%%%%%%%%%%%%%%%%%%%%%%%%%%%%%%%%%

\index{Ginzburg criterion}

The Ginzburg criterion explains the agreement between the values of the critical exponents above $d_c$ dimensions and the mean-field predictions.\cite{Ginzburg}
To discuss it, we return to the energy Eq.(\ref{DeltaH}) which was neglected in the mean-field Hamiltonian,
%\mycheck{\red{DeltaH2}}
\begin{equation}
 \Delta \mathcal{H} =
 - J\sum_{\langle{i,j}\rangle}{(s_i-m_L)( s_j - m_L)} 
= -J \sum_{\langle{i,j}\rangle}{G_L(r_i,r_j)} = - \frac{Jq}{2\beta}\chi_L,
\label{DeltaH2}
\end{equation}
having used the fluctuation-dissipation theorem. 
The neglect of this term is justified if its contribution to the energy  is small compared to that of the mean-field part.
From Eq.(\ref{III3i}), this is the case in zero field if $|\Delta \mathcal{H} | \ll \mathcal{H}_{\rm{MF}} $ or
%\mycheck{\red{Ginz2}}
\begin{equation}
 \frac{\chi_L}{\beta} \ll N m_L^2.
\label{Ginz2}
\end{equation}
In the infinite-volume limit, the procedure is to replace $N$ by $\xi^d$, so that the Ginzburg criterion becomes
%\mycheck{\red{Ginz3}}
\begin{equation}
 \frac{\chi_\infty}{\beta} \ll \xi_\infty^d m_\infty^2,
\label{Ginz3}
\end{equation}
or $\nu d > 2\beta + \gamma = 2-\alpha$. Using, for self-consistency, the mean-field values of the critical exponents this gives the criterion that $d>4$.
Therefore, mean-field and Landau theory should deliver meaningful results above $d_c=4$ dimensions.

We will revisit the Ginzburg criterion in Section~\ref{Conclusions}, in the light of developments outlined in the interim.

%%%%%%%%%%%%%%%%%%%%%%%%%%%%%%%%%%%%%%%%%%%%%%%%%%%%%%%%%%%%%%%%%%%
%\mycheck{\red{secwilsonRG}}
\section{Wilson's Renormalization-Group Theory}
\label{secwilsonRG}
%%%%%%%%%%%%%%%%%%%%%%%%%%%%%%%%%%%%%%%%%%%%%%%%%%%%%%%%%%%%%%%%%%%

\index{renormalization group}

Wilson's approach\cite{Wilson} goes beyond Kadanoff's heuristic approach in that new coupling constants can be generated with each renormalization-group transformation.
It delivers an explanation for universality and the calculation of critical exponents.
We require that the partition function $Z_L({\cal{\mathcal{H}}})$ be unchanged under the RG transformation, s.t. $  Z_{L^\prime} ({\mathcal{H}}^\prime)  = Z_{L} ({\mathcal{H}})$, 
where $L^\prime = b^{-1}L$. 
In terms of the free energy, this means
%\mycheck{\red{Y8.4}}
\begin{equation}
 f_{L^\prime} ({\mathcal{H}}^\prime)
 =
 b^d f_L ({\cal{H}})
\,.
\label{Y8.4}
\end{equation}
Lengths are reduced by a factor $b$ through the RG, and the spins are also rescaled
%\mycheck{\red{Y8.7}}
\begin{equation}
 s_x \rightarrow s^\prime_{x^\prime}
 = b^{d_\phi} s_x \quad {\mbox{or}} \quad \phi (z)  \rightarrow  \phi^\prime (z^\prime) = b^{d_\phi} \phi (z).
\label{Y8.7}
\end{equation}
The Hamiltonian is generally written 
%\mycheck{\red{Y8.12}}
\begin{equation}
 {\cal{H}} = {\vec{\mu}} {\vec{S}} =  \sum_i \mu_i S_i
,
\label{Y8.12}
\end{equation}
where $\vec{\mu}$ is a vector in the space of all possible parameters which may govern the system and where
${\vec{S}}$ represent different interactions. In the Ising or $\phi^4$ case, 
we may consider $\mu_1 = t$, $\mu_2 = h$, $\mu_3 = u$, etc., with 
$S_1=\int{d^d{\mathbf{r}}\ \,{ \phi^2({\mathbf{r}})  }/2}
$
$S_2=\int{d^d{\mathbf{r}}\ \,{ \phi^4({\mathbf{r}}) }/4}
$
and
$S_3=-\int{d^d{\mathbf{r}}\ \,{  \phi({\mathbf{r}})}}.
$

The transformation maps
%\mycheck{\red{Y8.13}}
\begin{equation}
 \vec{\mu} \rightarrow \vec{\mu}^\prime = R_b \mu
,
\label{Y8.13}
\end{equation}
 and fixed points are given by
%\mycheck{\red{Y8.14}}
\begin{equation}
  \vec{\mu}^* = R_b {\vec{\mu}}^* 
\,.
\label{Y8.14}
\end{equation}
Repeated application of the renormalization-group transformation reduces length scales by a factor of $b$. The correlation length  remains unchanged in the thermodynamic limit only if it is infinite or zero. 
We expand about a fixed point (\ref{Y8.14}), 
%\mycheck{\red{Y8.15,Y8.16}}
\begin{eqnarray}
\vec{\mu} & = & {\vec{\mu}}^* + \delta \vec{\mu}
,
\label{Y8.15}
\\
\vec{\mu}^\prime & = & {\vec{\mu}}^* + \delta \vec{\mu}^\prime
,
\label{Y8.16}
\end{eqnarray}
such that 
%\mycheck{\red{Y8.17}}
\begin{equation}
 \delta \vec{\mu}^\prime = A_b({\vec{\mu}}^*) \delta \vec{\mu} 
\,.
\label{Y8.17}
\end{equation}
Two successive scale transformations by factors $b_1$ and $b_2$ are assumed to be 
equivalent to a single transformation by a scale factor of $b_1b_2$, so that
%\mycheck{\red{Y*}}
\begin{equation}
 A_{b_1b_2} = A_{b_1} A_{b_2} 
\,.
\label{Y*}
\end{equation}
If the eigenvalues and eigenvectors of $A_b$ are given by
%\mycheck{\red{}}
\begin{equation}
 A_b \vec{v}_i = \lambda_i \vec{v}_i
,
\end{equation}
then Eq.(\ref{Y*}) gives that the $\lambda_i$ are homogeneous functions of $b$:
%\mycheck{\red{Y8.19}}
\begin{equation}
 \lambda_i (b) = b^{y_i}
\,.
\label{Y8.19}
\end{equation}
Expanding $\vec{\mu}$ near ${\vec{\mu}}^*$ in terms of the eigenvectors $ \vec{v}_i$, one has
%\mycheck{\red{Y8.20}}
\begin{equation}
 \vec{\mu} = {\vec{\mu}}^* + \sum_i g_i \vec{v}_i
\,.
\label{Y8.20}
\end{equation}
The $g_i$ here are called linear scaling fields. 
With $\delta \vec{\mu} = \sum_i g_i \vec{v}_i$, one now has 
%\mycheck{\red{}}
\begin{equation}
 \delta \vec{\mu}^\prime = \sum_i g_i^\prime \vec{v}_i = \sum_i g_i \lambda_i \vec{v}_i 
\,
\end{equation}
so that
%\mycheck{\red{Y8.22}}
\begin{equation}
 g_i^\prime = b^{y_i} g_i
\,.
\label{Y8.22}
\end{equation}
This gives how the linear scaling fields transform under the renormalization group and backs up the Kadanoff scaling picture.

If $\lambda_i>0$, the scaling field $g_i$ is augmented under renormalization group  and the system is driven away from the fixed point.
In this case, $\lambda_i$ is called a relevant scaling variable.
If $\lambda_i < 0$, the associated $g_i$ is an irrelevant scaling field and $\lambda_i$ is called an irrelevant scaling variable.
If $\lambda_i = 0$, it is marginal. 

We re-express Eq.(\ref{Y8.4}) in terms of the linear scaling fields as
%\mycheck{\red{}}
\begin{equation}
 f_\infty(\vec{\mu}) = b^{-d} f_\infty(\vec{\mu}^\prime)
\,.
\label{Y8.26}
\end{equation}
Near a fixed point, then, where $\vec{\mu}$ and $\vec{\mu}^\prime$ may be written in terms of  linear scaling fields,
this may again be rewritten as
%\mycheck{\red{Y8.27}}
\begin{equation}
 f_\infty(g_1,g_2,g_3, \dots)  = b^{-d} f_\infty\left({b^{y_1}g_1, b^{y_2}g_2, b^{y_3}g_3, \dots }\right)
\,.
\label{Y8.27}
\end{equation}
For the $\phi^4$ theory, the renormalization-group approach leads to two fixed points.
The Wilson-Fisher fixed point is stable below $d_c$ and the Gaussian fixed point is stable above $d_c$. 
 Eq.(\ref{Y8.22}) reads
\begin{eqnarray}
 r_0 & \rightarrow  & r_0^\prime = b^{y_t}r_0 
,\\ 
 h & \rightarrow & h^\prime = b^{y_h} h 
,\\
 u & \rightarrow & u^\prime = b^{y_u} u\,.
\end{eqnarray}
The first of these may also be written $ \rightarrow   t^\prime = b^{y_t} t$, after Eq.(\ref{ritot}). 
Eq.(\ref{Y8.27}) for the $\phi^4$ theory is then
%\mycheck{\red{Y8.28}}
\begin{equation}
 f_\infty(t,h,u) = b^{-d} f_\infty\left({b^{y_t}t, b^{y_h}h, b^{y_u}u }\right)
\,.
\label{Y8.28}
\end{equation}
This is the {\emph{scaling hypothesis}} (\ref{CM2.227}) with the irrelevant field accounted for.

Differentiating Eq.(\ref{Y8.28})  appropriately to obtain the thermodynamic functions, one finds
%\mycheck{\red{noDIR2}}
\begin{equation}
 \alpha = 2 - \frac{d}{y_t},\quad
 \beta = \frac{d-y_h}{y_t}, \quad
 \frac{1}{\delta} = \frac{d}{y_h}-1, \quad
 \gamma = \frac{2y_h-d}{y_t}.
\label{noDIR2}
\end{equation}
A similar form for the correlation function,
%\mycheck{\red{noDIR4}}
\begin{equation}
 G_\infty (r,t,h,u) = b^{-2d_\phi} G_\infty(b^{-1}r,b^{y_t}t,b^{y_h}h,b^{y_u}u),
\label{noDIR4}
\end{equation}
in which $d_\phi$ is the scaling dimension of the fields defined in Eq(\ref{Y8.7}),
delivers
%\mycheck{\red{noDIR5}}
\begin{equation}
 G_\infty (r,t,0,0) = b^{-2d_\phi} G_\infty(b^{-1}r,b^{y_t}t,0,0),
\label{noDIR5}
\end{equation}
having set both $h$ and $u$ to zero.
Then setting $b=x$, one finds
%\mycheck{\red{noDIR6}}
\begin{equation}
 G_\infty (r,t,0,0) = \frac{g(r/t^{-\frac{1}{y_t}})}{r^{2d_\phi}},
\label{noDIR6}
\end{equation}
for some function $g$. Comparing to the general form 
$ G_\infty (r,t) = \exp{-({{r}/{\xi}})}/{r^{d-2+\eta}}$,
one concludes
%\mycheck{\red{noDIR8,noDIR9}}
\begin{eqnarray}
 \nu & = & \frac{1}{y_t},
 \label{noDIR8}\\
 \eta & = & 2d_\phi+2-d.
 \label{noDIR9}
 \end{eqnarray}
These recover identities established in Kadanoff's approach (see Sec.~\ref{11.05.2009c}).
Alternatively, the scaling form
%\mycheck{\red{noDIR10}}
\begin{equation}
 \xi_\infty (t,h,u) = b \xi_\infty(b^{y_t}t,b^{y_h}h,b^{y_u}u).
\label{noDIR10}
\end{equation}
directly delivers Eq.(\ref{noDIR8}).

%%%%%%%%%%%%%%%%%%%%%%%%%%%%%%%%%%%%%%%%%%%%%%%%%%%%%%%%%%%%%%%%%%%%%%%%%%%
%\mycheck{\red{11.05.2009gaussian}}
\subsection{Scaling at the Gaussian fixed point}
\label{11.05.2009gaussian}
%%%%%%%%%%%%%%%%%%%%%%%%%%%%%%%%%%%%%%%%%%%%%%%%%%%%%%%%%%%%%%%%%%%%%%%%%%%

\index{fixed point!Gaussian}

At the Gaussian fixed point, the renormalization-group scaling dimensions are obtained by power counting:
%\mycheck{\red{DAIIdphi,DAIIdt,DAIIdu,DAIIdh}}
For the action (\ref{Amit6.7b}) to remain dimensionless under the transformation $r \rightarrow b^{-1}r$, $\phi \rightarrow b^{d_\phi}\phi$, $r_0 \rightarrow b^{y_t} r_0$, $u \rightarrow b^{y_u} u$, $h \rightarrow b^{y_h} h$, one requires
\begin{eqnarray}
% (\nabla \phi)^2 \rightarrow b^{2d_\phi-2} = b^ d  & \Rightarrow    & 
d_\phi & = & \frac{d}{2} - 1
,
\label{DAIIdphi}
\\ 
 y_t  & = &   2
,
\label{DAIIdt}
\\ 
   y_u    & = &   4-d
,
\label{DAIIdu}
\\
 y_h      & = & \frac{d}{2} + 1
\,.
\label{DAIIdh}
\end{eqnarray}

Inserting these into Eqs.(\ref{noDIR2}) and (\ref{noDIR8}), we identify
\begin{equation}
\alpha = 2 - \frac{d}{2}, \quad
\beta  = \frac{d}{4} - \frac{1}{2}, \quad
\gamma = 1, \quad
\frac{1}{\delta} = \frac{d-2}{d+2}, \quad
\nu = \frac{1}{2},
\label{noDIR11}
\end{equation}
with $\eta = 0$ from Eq.(\ref{noDIR9})

According to RG theory, the  Gaussian fixed point is stable at and above $d_c$, so these values are supposed to be valid there.
Therefore they should coincide with the results of Landau theory, 
$\alpha = 0$, $\beta = 1/2$, $\gamma = 1$, $\delta = 1/3$, $\nu = 1/2$ and $\eta = 0$.
We see that, while $\gamma$, $\nu$ and $\eta$ are in agreement, the values for $\alpha$, $\beta$ and $\delta$ disagree, except at $d=d_c$ itself.

%%%%%%%%%%%%%%%%%%%%%%%%%%%%%%%%%%%%%%%%%%%%%%%%%%%%%%%%%%%%%%%%%%%
%\mycheck{\red{DIV}}
\section{Dangerous Irrelevant Variables}
\label{DIV}
\setcounter{equation}{0}
%%%%%%%%%%%%%%%%%%%%%%%%%%%%%%%%%%%%%%%%%%%%%%%%%%%%%%%%%%%%%%%%%%%

\index{dangerous irrelevant variables}

To repair the shortcomings identified above the upper critical dimension, Fisher introduced the notion of dangerous irrelevant variables \cite{FiHa83}.
The danger should apply to the free energy because it is associated with  $\alpha$, $\beta$ and $\delta$.
However, since the values of $\nu$ and $\eta$ coming from the Gaussian fixed point are correct (they coincide with Landau theory), one does not expect danger for either the correlation function or the correlation length. 
Moreover, since the susceptibility is linked to the correlation function by the fluctuation-disappation theorem, one expects no danger for $\chi$ either.
Indeed, the value $\gamma = 1$ coming from the Gaussian fixed point is the same as that from mean-field theory.

%%%%%%%%%%%%%%%%%%%%%%%%%%%%%%%%%%%%%%%%%%%%%%%%%%%%%%%%%%%%%%%%%%%
%\mycheck{\red{DIVaaa}}
\subsection{The Thermodynamic Limit}
\label{DIVaaa}
%%%%%%%%%%%%%%%%%%%%%%%%%%%%%%%%%%%%%%%%%%%%%%%%%%%%%%%%%%%%%%%%%%%

Eq.(\ref{cMF}) shows that the mean-field specific heat behaves as $u^{-1}$ (in the broken symmetry phase).
Therefore the naive process of setting $u$ to zero, or ignoring its role in the free energy derivatives, is incorrect.
In fact, the second $t$-derivative $f_{tt} \left({x,y,z }\right)$ should scale as $ z^{-1}$ for small values of the third argument. 
This identifies the variable $u$ as dangerous in the $\phi^4$ theory; it cannot be set to zero.
From Eq.(\ref{cMF}), then, one expects
%\mycheck{\red{Y8.32b}}
\begin{equation}
 c_\infty(t,h,u) \sim b^{-d+2y_t-y_u } u^{-1} {\tilde{f}}_{tt} \left({b^{y_t}t }\right).
\label{Y8.32b}
\end{equation}
 Now setting $b = |t|^{-1/y_t}$, one has 
%\mycheck{\red{Y8.32c}}
\begin{equation}
 c_\infty(t,0,u) \sim |t|^{-\frac{d-2y_t+y_u}{y_t} }  {\tilde{f}}_{tt} \left({\pm1 }\right)
.
\label{Y8.32c}
\end{equation}
Similarly, Eq.(\ref{MFM}) gives that the mean-field spontaneous magnetisation behaves with $u$ as $u^{-1/2}$.
Therefore, in the mean-field case the first $h$-derivative $f_h\left({x,y,z}\right) $ behaves as $z^{-1/2}$ for small $z$,
so that $z$ cannot simply be set to $0$.
Instead,
%\mycheck{\red{mscc}}
\begin{equation}
 m_\infty(t,0,u) \sim   \sim b^{-d+y_h-\frac{1}{2}y_u} u^{-\frac{1}{2}} {\tilde{f}}_h\left({b^{y_t}t}\right) 
.
\label{mscc}
\end{equation}
Again setting $b = t^{-1/y_t}$, one obtains
%\mycheck{\red{Y8.33c}}
\begin{equation}
m_\infty(t,0,u)
\sim  t^{\frac{d-y_h+\frac{1}{2}y_u}{y_t}} 
.
\label{Y8.33c}
\end{equation}
For the critical isotherm, the role of the dangerous irrelevant variable is apparent from Eq.(\ref{deltais3}), 
which indicates that $f_h\left({0,y,z}\right) $ behaves as $z^{-1/3}$. Then  
%\mycheck{\red{msc2}}
\begin{equation}
 m_\infty(0,h,u)   \sim b^{-d+y_h-\frac{1}{3}y_u} {\tilde{f}}_h\left({, b^{y_h}h, b^{y_u}u}\right) 
,
\label{msc2}
\end{equation}
which leads to 
%\mycheck{\red{Y8.35c}}
\begin{equation}
m_\infty(0,h,u)
\sim  h^{\frac{d-y_h+\frac{1}{3}y_u}{y_h}}
.
\label{Y8.35c}
\end{equation}
Finally, since from Eq.(\ref{ggg}) the leading mean-field susceptibility does not depend on $u$, we may
infer that $u$ is not dangerous for $\chi$, as noticed above. 
Therefore Eq.(\ref{ggg}) is expected to hold.
Similar statements hold for the correlation function and correlation length.

From these considerations, one identifies
%\mycheck{\red{MFalpha,MFbeta,MFgamma,MFdelta}}
\begin{eqnarray}
 \alpha & = & -\frac{d-2y_t+y_u}{y_t} 
,
\label{MFalpha}
\\
\beta & = & \frac{d-y_h+\frac{1}{2}y_u}{y_t} 
,
\label{MFbeta}
\\
\gamma & = & \frac{2y_h-d}{y_t} 
,
\label{MFgamma}
\\
\frac{1}{\delta} &= & \frac{d-y_h+y_u/3}{y_h}
.
\label{MFdelta}
\end{eqnarray}
Eqs.(\ref{noDIR8}) and (\ref{noDIR9}) are expected to remain valid for $\nu$ and $\eta$ since neither the correlation length nor the correlation function are expected to be affected by the danger of $u$.
Finally, with the scaling dimensions (\ref{DAIIdphi})--(\ref{DAIIdt}) from power counting, one obtains the correct exponents
%\mycheck{\red{MFeltaa}}
\begin{equation}
 \alpha = 0,
 \quad 
 \beta  =  \frac{1}{2} , \quad 
 \gamma  =  1, \quad 
\delta =  3,
\label{MFdeltaa}
\end{equation}
along with
%\mycheck{\red{MFnua}}
\begin{equation}
\nu  =   \frac{1}{2}, \quad 
\eta =  0.
\label{MFnua}
\end{equation}

% % % % % % % % % % % % % % % % % % % % % % % % % % % % % % % % % % %
%\mycheck{\red{DIVFSS}}
\subsection{Finite-size scaling: Naive approach above $d_c$}
\label{DIVFSS}
% % % % % % % % % % % % % % % % % % % % % % % % % % % % % % % % % % %

Having using dangerous irrelevant variables to repair scaling in the thermodynamic limit above the upper critical dimension, we now turn to FSS there. 
A naive application of the FSS ansatz (\ref{II31}) replaces $\xi_\infty(t)$ by $L$ and one finds
%\mycheck{\red{LFSS}}
\begin{equation}
 c_L \sim L^{0}, \quad m_L\sim L^{-1}, \quad \chi_L \sim L^2, \quad \xi_L \sim L.
 \label{LFSS}
\end{equation}
We refer to this {\emph{Gaussian FSS}} or {\emph{Landau FSS}}  because it comes from applying the traditional FSS ansatz (\ref{II31}) to the mean-field exponents.  

Eq.(\ref{LFSS}) is, however in disagreement with an explicit analytical calculation by Br{\'{e}}zin for the $n$-vector model with periodic boundary conditions (PBCs).\cite{Br82} 
There have also been many numerical studies throughout the years \cite{Bi85,BNPY,RiNi94,Mon1996,PaRu96,LuBl97a,LuBl97b,BlLu97c,Luijtenthesis,BiLuReview,AkEr99,LuBi99,JoYo05,AkEr00,MeEr04,MeBa05,AkEr01,MeDu06} which confirm that Eqs.(\ref{LFSS})  for $m_L$ and for $\chi_L$ do not hold for the Ising model when  PBCs are used.
To understand how the conventional paradigm deals with this inconsistency, we turn to the Gaussian model.

%%%%%%%%%%%%%%%%%%%%%%%%%%%%%%%%%%%%%%%%%%%%%%%%%%%%%%%%%%%%%%%%%%%
%\mycheck{\red{GaussianFSS}}
\subsection{FSS in the Gaussian model with periodic boundaries}
\label{GaussianFSS}
%%%%%%%%%%%%%%%%%%%%%%%%%%%%%%%%%%%%%%%%%%%%%%%%%%%%%%%%%%%%%%%%%%%
\index{Gaussian model}

The Gaussian or free field theory has an action given by Eq.(\ref{Amit6.7b}) dropping the quartic self-interaction term. 
Defined on a finite-sized lattice in momentum space in vanishing field, it is given  as
%\mycheck{\red{Fri01}}
\begin{equation}
 S^{(0)}_L(\phi) = \frac{1}{2}\sum_k (k^2 + r_0^2)|\hat{\phi}_k|^2,
\label{Fri01}
\end{equation}
where $\hat{\phi}_k$ are the Fourier-transformed fields.

The correlation function is given through Eq.(\ref{LBpage26b}) by the inverse of the quadratic part of the action,
%\mycheck{\red{Fri03}}
\begin{equation}
 \tilde{G}^{(0)}(k) = \langle{\hat{\phi}_k \hat{\phi}_{-k}}\rangle
 =
 \frac{1}{k^2 + r_0}.
\label{Fri03}
\end{equation}
Here we have omitted the subscript $L$ from $G$, but it is understood that we are still considering a finite lattice.
Therefore, from the fluctuation dissipation  theorem,
%\mycheck{\red{LB1.4.8}}
\begin{equation}
 \chi_L = \tilde{G}^{(0)}(0) = r_0^{-1}.
\label{LB1.4.8}
\end{equation}
For PBCs,  the wave vectors are
%\mycheck{\red{Fri02}}
\begin{equation}
 k_\mu = \frac{2\pi n}{La}
\label{Fri02}
\end{equation}
in which  $ n = 1, 2, \dots L$ and $\mu = 1, \dots , d$.
 The correlation length can be defined as a second moment:
%\mycheck{\red{Fri5}}
\begin{equation}
 \xi_L = \frac{1}{k_{\rm{min}}} 
 \left[{
 \frac{1}{2d}
 \frac{\tilde{G}(0) - \tilde{G}(k_{\rm{min}})}{\tilde{G}(k_{\rm{min}})}
 }\right]^{\frac{1}{2}}
 = r_0^{-\frac{1}{2}},
\label{Fri5}
\end{equation}
where $k_{\rm{min}}= 2\pi / La$.
Therefore both the susceptibility and the correlation length diverge at $r_0=0$ in the Gaussian model, {\emph{even for a finite lattice}}.

This finite-size  divergence is problematic. To deal with it we examine the full $\phi^4$ action (\ref{Amit6.7b}), which in Fourier space is 
%\mycheck{\red{LB3.4.6M7.1Fri}}
\begin{equation}
S^{\rm{GLW}}_L[\phi] =  \frac{1}{2}
 \sum_{k}{
                    \left[{ 
                                    k^2 
                                    + r_0 
                    }\right] \hat{\phi}_k \hat{\phi}_{-k}
 }
+ \frac{u}{4} \frac{1}{L^d}
 \sum_{k_1,k_2,k_3 }{\hat{\phi}_{k_1} \hat{\phi}_{k_2}\hat{\phi}_{k_3}\hat{\phi}_{-k_1-k_2-k_3}}.
\label{LB3.4.6M7.1Fri}
\end{equation}
We gather the quadratic terms in the zero modes, 
%\mycheck{\red{Sat1}}
\begin{eqnarray}
{S}^{\rm{GLW}}_L[\phi] & =  & \frac{1}{2}
 \sum_{k}{
                    \left[{ 
                                    k^2 
                                    + r_0 
																		+ \frac{3u}{2L^d}\hat{\phi}_0^2
                    }\right] \hat{\phi}_k \hat{\phi}_{-k}
 }
+ \frac{u}{4L^d} \hat{\phi}_0^4
\nonumber \\
& & + \frac{u}{4} \frac{1}{L^d}
 {\sum^{}}^\prime_{k_1,k_2,k_3 }{\hat{\phi}_{k_1} \hat{\phi}_{k_2} \hat{\phi}_{k_3} \hat{\phi}_{-k_1-k_2-k_3}},
\label{Sat1}
\end{eqnarray}
where the prime indicates that the summation omits terms in which pairs of momenta vanish.
Again identifying the correlation function as the inverse of the quadratic part,
%\mycheck{\red{Sat4}}
\begin{equation}
 \tilde{G}^{(0)}(k) = \langle{\hat{\phi}_k \hat{\phi}_{-k}}\rangle
        = \frac{1}{r_0 + k^2 + \frac{3u}{2L^d}\langle{\hat{\phi}_0^2}\rangle}.
\label{Sat4}
\end{equation}
Therefore
%\mycheck{\red{Sat4b}}
\begin{equation}
\tilde{G}^{(0)}(0) = \frac{1}{r_0  + \frac{3u}{2L^d}\tilde{G}^{(0)}(0)}.
\label{Sat4b}
\end{equation}
Solving for $ \tilde{G}^{(0)}(0) $ when $r_0=0$, we obtain
%\mycheck{\red{Sat5}}
\begin{equation}
 \chi_L = G^{(0)}(0) = \langle{\hat{\phi}_0^2}\rangle =
\sqrt{\frac{2}{3u}}L^{\frac{d}{2}},
\label{Sat5}
\end{equation}
which is now finite.
This formula has been verified many times for finite-sized Ising systems with PBCs. \cite{Br82,Bi85,BNPY,RiNi94,Mon1996,PaRu96,LuBl97a,LuBl97b,BlLu97c,Luijtenthesis,BiLuReview,AkEr99,LuBi99,JoYo05,AkEr00,MeEr04,MeBa05,AkEr01,MeDu06,MeDu06}

The propagator is then
%\mycheck{\red{Sat6}}
\begin{equation}
 G^{(0)}(k) = \langle{\phi_k\phi_{-k}}\rangle
        = \frac{1}{r_0 + k^2 + \sqrt{\frac{3u}{2}}L^{-d/2}}
\label{Sat6}
\end{equation}
The correlation length is  
%\mycheck{\red{Sat7}}
\begin{equation}
 \xi_L^{(2)} = \frac{1}{k_{\rm{min}}} 
 \left[{
 \frac{1}{2d}
 \frac{\tilde{G}^{(0)}(0) - \tilde{G}^{(0)}(k_{\rm{min}})}{\tilde{G}^{(0)}(k_{\rm{min}})}
 }\right]^{\frac{1}{2}}
 = 
 \frac{1}{k_{\rm{min}}}
 \left[{
   \frac{1}{2d}\sqrt{\frac{2}{3u}} L^{\frac{d}{2}} k_{\rm{min}}^2
	 + \frac{1}{2d}-1
 }\right]^{\frac{1}{2}},
\label{Sat7}
\end{equation}
when $r_0=0$.
Now, ${k_{\rm{min}}} = 2\pi /La$ from Eq.(\ref{Fri02}). Therefore 
$\xi_L(0) \sim L(L^{d/2-2} + \rm{const.})^{1/2}$, so that
%\mycheck{\red{firstQ}}
\begin{equation} 
  \xi_L(0) \sim
\left\{ \begin{array}{ll}
                                                                L & \mbox{if $d > 4$ } \\
                                                                   ~ & ~ \\
                                                                 \frac{d}{4}  & \mbox{if $d>4$} \\
                                                  \end{array}
                                                       \right..
\label{firstQ}
\end{equation}
Therefore both $\xi_L$ and $\chi_L$ remain finite at $r_0=0$ ($T=T_c$).

The result $\xi_L \sim L^{d/4}$ was also obtained on a theoretical basis for the large-$n$ vector model in Ref.~[\refcite{Br82}].
However, it violates an expectation that the correlation length be bounded by the length.\cite{BNPY}
Nonetheless, it has been verified numerically for Ising systems with periodic boundaries in 
 Refs.~[\refcite{JoYo05,ourNPB,ourCMP}].
See also Refs.~[\refcite{YoungSG1,YoungSG2,Martin}] for studies of spin-glass systems.

We next move on to the general FSS scheme above the upper critical dimension with dangerous irrelevant variables. 
A question to keep in mind is whether the FSS behaviour outlined here is captured by the general scheme.
We will see it that, in its original formulation, it is not. 
This forced the introduction of another length scale (dubbed {\emph{thermodynamic length}) to control FSS above $d_c$.

%%%%%%%%%%%%%%%%%%%%%%%%%%%%%%%%%%%%%%%%%%%%%%%%%%%%%%%%%%%%%%%%%%%
%\mycheck{\red{secDIRS}}
\subsection{FSS with dangerous irrelevant variables}
\label{secDIRS}
%%%%%%%%%%%%%%%%%%%%%%%%%%%%%%%%%%%%%%%%%%%%%%%%%%%%%%%%%%%%%%%%%%%
\index{dangerous irrelevant variables}

In 1984, Binder, Nauenberg, Privman and Young extended Fisher's concept of dangerous
irrelevant variables to the finite-volume case \cite{BNPY}.
We follow and assume that the finite-size counterpart of Eq.(\ref{Y8.28}) is
%\mycheck{\red{BNPY1}}
\begin{equation}
 f_L(t,h,u) = b^{-d} f_{L/b}\left({ tb^{y_t}, hb^{y_h}, ub^{y_u}  }\right)
.
\label{BNPY1}
\end{equation}
We make the further assumption that, for small $x_3$, 
%\mycheck{\red{RG2}}
\begin{equation}
f_{L/b}(x_1,x_2,x_3) = x_3^{p_1}{\bar{f}}_{L/b} \left({x_1x_3^{p_2},x_2x_3^{p_3}}\right)
.
\label{RG2}
\end{equation}
Under this assumption, the free energy may be written
%\mycheck{\red{RG3}}
\begin{equation}
 f_L(t,h,u) = b^{-d^*} {\bar{f}}_{L/b}\left({ tL^{y_t^*}, hL^{y_h^*} }\right)
,
\label{RG3}
\end{equation}
in which 
%\mycheck{\red{RG4,RG5,RG6}}
\begin{eqnarray}
d^* & = & d-p_1 y_u
,
\label{RG4}\\
y_t^* & = & y_t+p_2y_u
,
\label{RG5}\\
y_h^* & = & y_h+p_3y_u.
\label{RG6}
\end{eqnarray}
Here  $y_t^* $ and $y_h^* $  effective exponents. 
The infinite volume limit of Eq.(\ref{RG3}) yields
%\mycheck{\red{RG8}}
\begin{equation}
 f_\infty(t,h) =  t^{\frac{d^*}{y_t^*} }
   Y_{\pm}\left({  ht^{-\Delta} }\right)
,
\label{RG8}
\end{equation}
in which 
%\mycheck{\red{RG9}}
\begin{equation}
 \Delta = \frac{y_h^*}{y_t^*}.
\label{RG9}
\end{equation}
Now we take derivatives to obtain the thermodynamic functions.
These give
%\mycheck{\red{RG10,RG11,RG12,RG122}}
\begin{eqnarray}
\alpha & = & 2 - \frac{d^*}{y_t^*}
, \label{RG10} 
\\
\beta & = &  \frac{d^* - y_h^*}{y_t^*}
, \label{RG11} 
\\
\gamma & = & \frac{2y_h^* - d^* }{y_t^*}
, \label{RG12} \\
\frac{1}{\delta} & = & \frac{d^*}{y_h^*} - 1 
\label{RG122} 
,
\end{eqnarray}
which in turn yield
%\mycheck{\red{BNPY9,BNPY10,RG13}}
\begin{eqnarray}
y_t^* & = & \frac{d^*}{2-\alpha} = \frac{d^*}{\gamma + 2\beta}
, \label{BNPY9} 
\\
y_h^* & = & \frac{d^* \delta}{1 + \delta} =\frac{d^*(\gamma + \beta)}{\gamma + 2\beta}
, \label{BNPY10} 
\\
\Delta & = & \beta + \gamma 
\label{RG13} 
.
\end{eqnarray}
The last of these establishes $\Delta$ as the gap exponent.
Also, it is at this point that we obtain the static scaling relations
\[
 \alpha + 2 \beta + \gamma = 2, \quad {\mbox{and}} \quad \beta(\delta -1) = \gamma.
\]
Inserting the mean-field values $\alpha = 0$, $\beta = 1/2$, $\gamma = 1$ and $\delta = 1/3$, we obtain
%\mycheck{\red{}}
\begin{equation}
 y_t^* = \frac{d^*}{2},
 \quad
 \mbox{and}
 \quad
 y_h^* = \frac{3d^*}{4}.
\end{equation}
In Ref.~[\refcite{BNPY}] three arguments are given for the coincidence of $d^*$ and $d$. 
This equality corresponds to $p_1=0$.
(We revisit this in Sec.~\ref{thermolength}.)
In this case, one has
%\mycheck{\red{dfg1}}
\begin{equation}
 y_t^* = \frac{d}{2},
 \quad
 \mbox{and}
 \quad
 y_h^* = \frac{3d}{4}.
\label{dfg1}
\end{equation}
Moreover,
%\mycheck{\red{dfg11}}
\begin{equation}
 p_2 = -\frac{1}{2}
 \quad
 \mbox{and}
 \quad
 p_3 = -\frac{1}{4}.
\label{dfg11}
\end{equation}
We can now determine the FSS of the thermodynamic functions by appropriate differentiation of Eq.(\ref{RG3}).
Differentiating with respect to $h$, setting $t=h=0$ and $b=L$, one obtains
$m_L(0,0) \sim L^{-d+y_h^*}$ and $\chi_L(0,0) \sim L^{-d+2y_h^*}$.
From Eq.(\ref{dfg1}) then, 
%\mycheck{\red{dfg2}}
\begin{equation}
 m_L \sim L^{-\frac{d}{4}},
 \quad
 \mbox{and}
 \quad
 \chi_L \sim L^{\frac{d}{2}}.
\label{dfg2}
\end{equation}
Similarly differentiating Eq.(\ref{RG3}) with respect to $t$ one obtains 
$c_L(0,0) \sim L^{-d+2y_t^*} $ or
%\mycheck{\red{dfg3}}
\begin{equation}
 c_L \sim L^0.
\label{dfg3}
\end{equation}

Eqs.(\ref{dfg2}) are different from the naive Landau FSS results of Eqs.(\ref{LFSS}).
Eqs.(\ref{dfg2}) are, however, in agreement with the result from the Gaussian model (\ref{Sat5}) with PBCs.
Indeed, many numerical studies using PBCs throughout the years have verified that Eqs.~(\ref{dfg2})  give the correct FSS above the upper critical dimension.
However, widespread belief is that Landau FSS holds for free boundary conditions (FBCs).
This is still a subject of debate and we defer discussion until Section~\ref{Conclusions}.

In Ref.~[\refcite{BNPY}], an extension of the above scenario to the correlation length or correlation function was not fully considered. 
The equivalent form to Eq.(\ref{BNPY1}) for the correlation length is
\begin{equation}
 \xi_L(t,h,u) = b \xi_{L/b}\left({ tb^{y_t}, hb^{y_h}, ub^{y_u}  }\right)
.
\label{RG14}
\end{equation}

Following a similar procedure to before, we write
\begin{equation}
 \xi_L(t,h,u) = b^{1+q_1y_u} {\bar{\xi}}_{L/b}\left({ tb^{y_t+q_2y_u}, hb^{y_h+q_3y_u} }\right)
,
\label{RG16a}
\end{equation}
to obtain
%\mycheck{\red{RG16}}
\begin{equation}
 \xi_L(t,h) = L^{q} {\bar{\xi}}_{1}\left({ tL^{y_t^{**}}, hL^{y_h^{**}} }\right)
,
\label{RG16}
\end{equation}
in which 
%\mycheck{\red{RG17,RG18,RG19}}
\begin{eqnarray}
q & = & 1+q_1 y_u ,\label{RG17}
\\
y_t^{**} & = & y_t+q_2y_u, \label{RG18}
\\
y_h^{**} & = & y_h+q_3y_u. \label{RG19}
\end{eqnarray}
Setting $t=h=0$ gives a correlation length which scales algebraically with $L$.
However, believing that the ``finite-size correlation length $\xi_L$ is bounded by the length $L$'', in Ref.~[\refcite{BNPY}] it was assumed  $q_1=0$ so that $\xi_L \sim L$ for periodic as well as free boundaries.

The FSS prescription (\ref{II31}) therefore fails above $d=d_c=4$. 
This led Binder to consider alternatives and in 1985 he introduced the {\emph{thermodynamic length}}.\cite{Bi85}
Our next step is to summarise these arguments.

%%%%%%%%%%%%%%%%%%%%%%%%%%%%%%%%%%%%%%%%%%%%%%%%%%%%%%%%%%%%%%%%%%%
%\mycheck{\red{thermolength}}
\subsection{The thermodynamic length}
\label{thermolength}
%%%%%%%%%%%%%%%%%%%%%%%%%%%%%%%%%%%%%%%%%%%%%%%%%%%%%%%%%%%%%%%%%%%

\index{thermodynamic length}

In order to repair FSS, Binder introduced the concept of {\emph{thermodynamic length}}.
In the $T<T_c$ phase, the probability density of the magnetisation $M$ for a finite system may be approximated by the sum of two Gaussians \cite{Schladming}.
The Gaussians are centred around the (infinite-volume) spontaneous magnetisation 
$\langle{M}\rangle= Nm_{\rm{sp}}=Nm_\infty(0,0)$ and are of width 
$\sqrt{\langle{(M-\langle{M}\rangle)^2}\rangle} = N\chi_\infty(t,0)/\beta$.
The  arguments of the Gaussians are therefore 
\[
 \frac{N(m \pm m_{\rm{sp}})^2}{2k_BT\chi_\infty}
=
 \frac{m_{\rm{sp}}^2(1 \pm m/m_{\rm{sp}})^2}{2k_BT\chi_\infty}L^d
=
 \frac{1}{2k_BT}
 \left({
        1 \pm \frac{m}{m_{\rm{sp}}}
				}\right)^2
 \left({
        \frac{L}{\ell_\infty}
				}\right)^d,
\]
in which
%\mycheck{\red{ell}}
\begin{equation}
 \ell_\infty = \left({
                       \frac{\chi_\infty}{m_{\rm{sp}}^2}
							       }\right)^{\frac{1}{d}}.
\label{ell}
\end{equation}
Now, with $\chi_\infty(t) \sim |t|^{-1}$ and  $m_\infty(t) \sim |t|^{1/2}$, we have
%\mycheck{\red{ellsc}}
\begin{equation}
 \ell_\infty(t) \sim |t|^{-\frac{2}{d}}.
\label{ellsc}
\end{equation}
This is called the {\emph{thermodynamic length}} because it appears in the thermodynamic functions.

The assumption is that  $\ell_\infty(t)$ governs FSS instead of $\xi_\infty(t)$ (which scales as $t^{-1/2}$.
Indeed,  the FSS ansatz (\ref{II31})  is then replaced by 
%\mycheck{\red{tlFSS}}
\begin{equation}
 \frac{P_L(t)}{P_\infty(t)}
 =
 {\mathcal{F}}_P
 \left[{\frac{L}{\ell_\infty(t)}}\right].
 \label{tlFSS}
\end{equation}
Applied to the magnetisation, for example, this ansatz gives
\[
 m_L(0) = m_\infty(t) {\mathcal{F}}_m
 \left({\frac{L}{t^{-2/d}}}\right) \sim L^{-\frac{d}{4}},
\]
in agreement with Eq.(\ref{dfg2}).
We can check that the new ansatz also delivers FSS for the susceptibility  in Eq.(\ref{dfg2}).

This set-up is in accordance with the change in the homogeneity assumption 
from $tb^{y_t}$  in Eq.(\ref{BNPY1}) to the combination  
$tb^{y_t^*} = [b/t^{-1/y_t^*}]^{y_t^*}
= [b/\ell_\infty(t)]^{y_t^*}$ provided that $\ell_\infty (t) \sim |t|^{-1/y_t^*}$.
Comparing with Eq.(\ref{ellsc}), one has $y_t^*=d/2$.
This  justifies the identification of $d^*$ with $d$ leading to Eq.(\ref{dfg1}).

One notes that this profoundly modifies Fisher's original (infinite-volume) dangerous-irrelevant-variables mechanism of Sec.~\ref{DIV} because, not only the prefactor of the scaling function is altered in the $u \rightarrow 0$ limit, but also its argument.

The finite-size counterpart of the thermodynamic length $\ell_\infty$ was  termed {\emph{coherence length}} $\ell_L$ in Ref.~[\refcite{BDT}] and it scales as the system extent $L$. 
A so-called {\emph{characteristic length}} $\lambda_L(t)$ was also introduced, as the FSS counterpart of the infinite-volume correlation length $\xi_\infty$. 

The picture set out above was, until recently, essentially the basis for the standard understanding of scaling and FSS above the upper critical dimension. 
The thermodynamic length $\ell_\infty(t)$ is supposed to replace $\xi_\infty(t)$ in the FSS scaling ansatz (\ref{II31}), so that 
%\mycheck{\red{II31l}}
\begin{equation}
 \frac{P_L(t)}{P_\infty(t)}
 =
 {\mathcal{F}}_P
 \left[{\frac{L}{\ell_\infty(t)}}\right].
 \label{II31l}
\end{equation}

Although all of this delivers the correct values for the exponents above $d_c$, it is not satisfactory.
Along with the lattice spacing $a$ this means a number of length scales have entered the game.
There have been many instances of proliferation in science which signalled a flaw in a standing paradigm and we are reminded of the quote in the Introduction. 
We next outline more concrete reasons for dissatisfaction before introducing the new picture.

%%%%%%%%%%%%%%%%%%%%%%%%%%%%%%%%%%%%%%%%%%%%%%%%%%%%%%%%%%%%%%%%%%%
%\mycheck{\red{shortcomings}}
\section{An unsatisfactory paradigm}
\label{shortcomings}
\setcounter{equation}{0}
%%%%%%%%%%%%%%%%%%%%%%%%%%%%%%%%%%%%%%%%%%%%%%%%%%%%%%%%%%%%%%%%%%%

%The paradigm, from the above account, for scaling and FSS above the upper critical dimension is as follows.

Dangerous irrelevant variables play a crucial role in reconciling scaling above the upper critical dimension with mean-field and Landau theory in the thermodynamic limit, where they alter the prefactor 
of some of the scaling functions. 
A naive approach to FSS then delivers Gaussian or Landau FSS in which   $\chi_L \sim L^{\gamma / \nu} = L^2$ and $\xi_L \sim L$.
These are incompatible, however, with exact calculations in the Gaussian model, the $n$-vector model and with the results of numerical simulations with PBCs. 
To repair this, a modification of the role of dangerous irrelevant variables was introduced whereby they also alter the arguments of the scaling functions for finite-size systems. 
This requires the introduction of a new length scale -- the thermodynamic length -- which takes over from the correlation length in the finite-size scaling ansatz. 

The resulting predictions for $m_L$, $\chi_L$ and $c_L$ given by Eqs.(\ref{dfg2}) and (\ref{dfg3}) have been confirmed many times over using numerical simulations of the Ising model with PBCs  in 
%four \cite{Luijtenthesis,JoYo05,KeLa91,Ak01,KeLa94,Ke04,KJJ2006}, 
five \cite{Bi85,BNPY,RiNi94,Mon1996,PaRu96,LuBl97a,BlLu97c,Luijtenthesis,BiLuReview,AkEr99,LuBi99,JoYo05}, 
six \cite{AkEr00,MeEr04,MeBa05}, 
seven \cite{AkEr01,MeDu06} and eight \cite{MeDu06} dimensions.

In contrast to the  PBC case, however, there have  been  few studies of systems with FBCs above $d_c$ \cite{ourNPB,ourCMP,RuGa85,LuMa11}.
%  which are complicated by additional scaling fields associated with boundaries in the RG picture \cite{BDT,Bi83Diehl86}.
The standard picture is that the FBC case is governed by  Gaussian FSS, in which $\chi_L \sim L^2$
due to a belief that Eq.(\ref{dfg2}) ``cannot hold for FBCs because it lies above a strict upper bound''\cite{RuGa85} (namely $L^{\gamma/\nu}=L^2$) established in Ref.~[\refcite{Watson}] (see also Ref.~[\refcite{Gunton}]). 
A recent numerical study of the five-dimensional Ising model with FBC's appears to 
support Gaussian FSS  $\chi_L \sim L^{2}$ at the critical  point $T_c$.

However, a number of unsettling issues with the standard picture arise.

Firstly, the  Fourier analysis of Ref.~[\refcite{RuGa85}], which yielded Landau FSS for FBC's, neglects the quartic part of the $\phi^4$ action. This omission was justified by an expectation that the Gaussian result should apply to leading order. 
However, it was shown in Refs.~[\refcite{LuBl97a,LuBl97b,BlLu97c}] that the  FSS behaviour (\ref{dfg2})  is obtained from  precisely this interaction term in the PBC case.

Secondly, the mechanism outlined in Sec.~\ref{secDIRS} and the FSS ansatz (\ref{II31l}) do not explicitly 
distinguish between different sets of boundary conditions. So the origin of the disparity between FSS with PBC's and FBC's is unclear.

It also remains unexplained why the dangerous irrelevant variables mechanism affects the free energy but not the correlation function or the correlation length in the PBC case.
This is especially puzzling if the arguments of the thermodynamic functions are affected as well as the prefactors, because the same arguments enter into all three functions.

Br{\'{e}}zin \cite{Br82}  established that in the large-$n$ limit of the $n$-vector model the correlation length for the finite system scales as $\xi_L \sim L^{d/4}$ for $d>4$. In the $d=4$ case, the corresponding FSS is $\xi_L \sim L (\ln{L})^{1/4}$. He  argued for the same behaviour in the finite-$n$ case.
In Ref.~[\refcite{KeLa91}], an alternative ansatz to (\ref{II31}) was introduced to deal with the case of logarithmic corrections at the critical dimension itself:
%\mycheck{\red{II31k}}
\begin{equation}
 \frac{P_L(t)}{P_\infty(t)}
 =
 {\mathcal{F}}_P
 \left[{\frac{\xi_L(t)}{\xi_\infty(t)}}\right].
 \label{II31k}
\end{equation}
In the $d=4$ case, the scaling dimension $y_u$ vanishes and $u$ is marginal rather than irrelevant.
The ansatz (\ref{II31k}) is therefore not dependent on the danger of $u$.
For the Ising model or $\phi^4$ theory in $d=4$ dimensions, it gives the correct FSS for the thermodynamic functions and partition function zeros \cite{KeLa91,KeLa94, Ke04}, provided  
Eq.(\ref{corrLlog}) holds with Br{\'{e}}zin's prediction $\hat{\q} = 1/4$.
It also delivers the correct FSS above $d_c$ in the PBC case with $\xi_L \sim L^{d/4}$.
The ansatz (\ref{II31k}) reduces to  Eq.(\ref{II31}) in the case where $\xi_L \sim L$.
However, it remains distinct from Eq.(\ref{II31l}).
%The thermodynamic length is also supposed to coincide with the correlation length below $d_c$.

Thus the principle is established that the correlation length can, in fact, exceed the length of the system,
in contrast to the assumptions of Sec.~\ref{secDIRS} and of Ref.~[\refcite{BNPY}].
This was reaffirmed in Ref.~[\refcite{JoYo05}] for PBC's in a numerical simulation of the $d=4$ and $d=5$ Ising models. In particular, this confirmed the FSS of the correlation length as $\xi_L \sim L^{d/4}$.

 The less frequently  studied case of FBC's was  described in Ref.~[\refcite{JoYo05}] as ``poorly understood''.
Indeed, it is stated there that for FBC's ``it seems obvious that even for $d > 4$ the behavior of the system will be affected when $\xi_L$ becomes of order $L$'', rather than the larger $L^{d/4}$.  
On this basis, it was believed that the standard (Gaussian) FSS expressions (corresponding to $\xi_L \sim L$) are expected to apply. 
The larger length scale $L^{d/4}$ was then expected to contribute to corrections to scaling in some  manner which was unspecified.
The numerical results of Ref.~[\refcite{LuMa11}] for FBC's would at first sight  appear to substantiate these conclusions and speculations.

%%%%%%%%%%%%%%%%%%%%%%%%%%%%%%%%%%%%%%%%%%%%%%%%%%%%%%%%%%%%%%%%%%%
\subsection{Fisher's scaling relation for finite-size systems}
\label{fshortcomings}
%%%%%%%%%%%%%%%%%%%%%%%%%%%%%%%%%%%%%%%%%%%%%%%%%%%%%%%%%%%%%%%%%%%

\index{scaling relation! Fisher}

There is  another problem with the conventional paradigm.
To explain it, we return to the derivation of Fisher's scaling relation in Section~\ref{11.05.2009c}.
The expression equivalent to Eq.(\ref{Dissresp}) for a finite lattice system is
%\mycheck{Dissrespf}
\begin{equation} 
 \chi_L (t)  =  \int_a^{L}{dr\ \! r^{d-1} G_L(r,t)},
\label{Dissrespf}
\end{equation}
in which $a$ is the lattice spacing and we have dropped  explicit dependency on $h$ which is set to vanish.
Then, from the form (\ref{CM2.182}),
%\mycheck{Fisherreln1f}
\begin{equation}
 \chi_L = \int_0^{L}{dr\ \! r^{1-\eta} }
								       \mathcal{G}_{\pm} \left({\frac{r}{\xi_L}}\right),
\label{Fisherreln1f}											
\end{equation}
assuming that the lower integral limit in Eq.(\ref{Dissrespf}) only delivers corrections to scaling.
Fixing $r/\xi_L(t) = x$, this gives
%\mycheck{Fisherreln2f}
\begin{equation}
 \chi_L = \xi_L^{2-\eta} 
                  \int_{0}^{L/\xi_L}{dx\ \! x^{1-\eta} }
								       \mathcal{G}_{\pm} \left({x}\right).
\label{Fisherreln2f}		
\end{equation}
When $d<d_c$, where $\xi_L \sim L$, this delivers Fisher's scaling relation
\[
 \frac{\gamma}{\nu} = 2 - \eta.
\]

Above $d_c$, however, Eq.(\ref{Fisherreln2f}) runs into trouble.
According to the conventional  paradigm, $\chi_L \sim L^{d/2}$, at least for systems with PBC's.
With $\xi_L \sim L$, Eq.(\ref{Fisherreln2f}) leads to
\[
 \eta = 2 - \frac{d}{2}.
\]
This corresponds to a negative anomalous dimension above $d_c$, in disagreement with Landau theory, for which  $\eta = 0$.

The negativity of the  anomalous dimension 
was already noticed in a numerical study by Nagle and Bonner in Ref.~[\refcite{NaBo70}].
Baker and Golner also determined a negative $\eta$ in an analytical study of an Ising model for which scaling is exact \cite{BaGo73}. 
Their explanation was that there are two long-range distance scales: long long range and short long range.
They posited that long long-range order is controlled by a new, {\emph{different}} anomalous dimension, while  ``short long-range order'' is controlled by the usual $\eta$. 
The long long-range exponent  {\it{fails to satisfy the scaling relation for the anomalous dimension above the upper critical dimension.}}

Luijten and Bl{\"{o}}te revisited the problem using the relationship (\ref{II14}) to obtain the correlation function  from the free energy by differentiation with respect to two local fields, \cite{LuBl97b,Luijtenthesis}
\begin{equation}
 \langle{\phi(0)\phi(r)}\rangle \sim L^{-d} \frac{\partial^2 f(t,h,u)}{\partial h(0)\partial h(r)}.
\end{equation}
Using Eq.(\ref{BNPY1}) and ignoring the danger of $u$, one finds
$G_L \propto L^{2(y_h-d)} = L^{-(d-2)}$, which is the standard, Landau result with $\eta = 0$.
However, taking account of the dangerous irrelevancy by differentiating Eq.(\ref{RG3}) instead, 
Luijten and Bl{\"{o}}te obtained $G_L \propto L^{2(y_h^*-d)} = L^{-d/2}$, 
corresponding to an anomalous dimension  $\eta^* \equiv 2-d/2$.

The problem with this interpretation is that under the standard paradigm, although $u$ is dangerous for the free energy, it is not supposed to be dangerous for the correlation function.
The approach outlined here skirts the issue by appealing to the free energy rather than directly to the correlation function.

Luijten and Bl{\"{o}}te give a second interpretation to their anomalous dimensions \cite{LuBl97b,Luijtenthesis}. 
Eq.(\ref{Sat5}) gives the inverse correlation function at criticality in momentum space as
\[
G^{-1}(k) \sim  k^2 + \sqrt{\frac{3u}{2}} L^{-\frac{d}{2}}.
\]
If we identify $L^{-1} = k_{\rm{min}}$ as a momentum, we have two different decay modes.
We compare these to the general form $G^{-1}(q) = q^{2-\eta}$.
If we again ignore the danger and set $u=0$, we obtain $\eta = 0$.
When $k$ is large compared to $L^{-1} = k_{\rm{min}}$ (i.e., over distances significantly shorter than the lattice extent), we obtain the same result.
If, on the other hand, we acknowledge the danger and keep the $u$ term, it delivers an anomalous dimension $\eta^* = 2 - d/2$.

So the interpretation Refs.~[\refcite{LuBl97b,Luijtenthesis}] is that the short long-distance behaviour is governed by the exponent $\eta = 0$, which remains unaffected by the morphing of $y_h$ into $y_h^*$.
Long long-distance behaviour is then supposed to be ruled by $\eta^*$.

These interpretations, however, do not explain how $\eta^* = 2-d/2$ for long long distance is manifest as $\eta=0$ in the infinite-volume limit where  field-theoretic theorems outlawing negative anomalous dimensions apply \cite{Fi64sc,Fi69inequalities,Delamotte}.
Nor do they explain why it is the supposedly long long-range anomalous dimension $\eta^*$ associated 
with the correct dangerous-irrelevant-variables mechanism which conflicts with Landau and mean-field theory, fails to satisfy Fisher's relation and violates field theory.
One would rather expect the disparity to be associated with the incorrect neglect of dangerous irrelevant variables or taking short, rather than long,  long distances.
Therefore {\emph{the standard paradigm does not explain scaling above the upper critical dimension.}}

To summarise, the standard paradigm for scaling and FSS above the upper critical dimension is assembled in a rather ad hoc fashion. It posits that FSS becomes non-universal above $d_c$, where hyperscaling breaks down and extra length scales appear. Nonetheless, the numerical results were interpreted as in agreement with the detail of the theory and critical exponents are known.
Thus we return to the statement by Binder et al. in quoted in Sec.\ref{Introduction}\cite{Binderquote}:
``although \dots all exponents are known, \dots the existing theories clearly are not so good''.

% Mention spin glass - FOR PBC's ONLY!

%%%%%%%%%%%%%%%%%%%%%%%%%%%%%%%%%%%%%%%%%%%%%%%%%%%%%%%%%%%%%%%%%%%
%\mycheck{\red{longcomings}}
\section{A New Paradigm: Hyperscaling above the Upper Critical Dimension and a Negative Anomalous Dimension}
\label{longcomings}
\setcounter{equation}{0}
%%%%%%%%%%%%%%%%%%%%%%%%%%%%%%%%%%%%%%%%%%%%%%%%%%%%%%%%%%%%%%%%%%%

We next present a recently developed  picture, \cite{ourNPB,ourCMP,ourEPL} starting with a derivation which  illustrates the need for a new theory.

%%%%%%%%%%%%%%%%%%%%%%%%%%%%%%%%%%%%%%%%%%%%%%%%%%%%%%%%%%%%%%%%%%%
\subsection{Self-Consistency: the Requirement for a New Universal Scaling Exponent $\q$}
\label{longcomings2}
%%%%%%%%%%%%%%%%%%%%%%%%%%%%%%%%%%%%%%%%%%%%%%%%%%%%%%%%%%%%%%%%%%%

Here we explore the self-consistency of the FSS ans{\"{a}}tze (\ref{II31l}) and (\ref{II31k}).
Each reduces to the form (\ref{II31}) when $d<d_c$.

For finite Ising systems the Lee-Yang zeros form a discrete set on the imaginary-$h$ axis \cite{LY}. 
We label them $h_j(L,t)$, with $j$ indicating relative positions according to distance from the real-$h$ axis. 
Since the total number of zeros is proportional to the volume $L^d$, these are expected to scale as a function of the ratio $j/L^d$, at least for sufficiently large $j$ \cite{IPZ,JaKe01}.
(In the case of $d=d_c=4$ dimensions, logarithmic corrections do not enter this ratio.\cite{KJJ2006})
The counterpart of $h_1$ in the infinite-$L$ limit is the Lee-Yang edge $h_{\rm{YL}}(t)$.
Following Ref.~[\refcite{KJJ2006}],  the finite-size susceptibility is related to the Lee-Yang zeros as
%\mycheck{\red{chih}}
\begin{equation}
 \chi_L \sim  L^{-d} \sum_{j=1}^{L^d}{{h_j^{-2}(L)}}.
\label{chih}
\end{equation}
This holds {\emph{irrespective of boundary conditions}}. 

We first consider the standard paradigm and the ansatz  (\ref{II31l}) corresponding to the thermodynamic-length formalism, and $\ell_\infty(t) \sim  |t|^{-2/d}$ from Eq.(\ref{ellsc}).
The thermodynamic length, being defined for infinite volume, is not affected by boundary conditions and the ansatz gives $h_j(L) \sim  ({{j}/{L^d}})^{{\Delta }/{2}}$.
Eq.(\ref{chih}) is then 
\[
 \chi_L \sim L^{d(\Delta - 1)}
 \sum_{j=1}^{L^d}{j^{-\Delta}}.
\]
In the PBC case,  the left hand side of this equation is $\chi_L \sim  L^{d/2}$. Matching with the right hand side, one obtains $\Delta = 3/2$, which is, indeed, correct. 
However, if $\chi_L \sim  L^{2}$ (the supposed FBC case), one arrives at $\Delta = 1+d/2$.
This is incorrect, except when $d=4$.
Therefore the standard paradigm leads to an {\emph{inconsistency}} when the boundary conditions are free.

We next consider the ansatz (\ref{II31k}) and we allow
%\mycheck{\red{mainq}}
\begin{equation}
 \xi_L(0) \sim L^{\q}.
\label{mainq}
\end{equation}
The exponent $\q$ is the leading power-law analogue to $\hat{\q}$ of Eq.(\ref{corrLlog}), 
which is well established at least for PBC's.
For the $j$th zero and the magnetic susceptibility it gives
%\mycheck{\red{hjL}}
\begin{equation}
 h_j(L) \sim  \left({\frac{j}{L^d}}\right)^{\frac{\Delta \q}{\nu d}}
 \label{hjL}
\end{equation}
and
%\mycheck{\red{FSSchi}}
\begin{equation}
 \chi_L \sim  L^{\frac{\gamma \q}{\nu }},
 \label{FSSchi}
\end{equation}
respectively.
% Gaussian or Landau FSS is recovered simply by setting $\q=1$.
Eq.(\ref{chih}) then gives
%\mycheck{\red{21}}
\begin{equation}
 L^{\frac{\gamma \q}{\nu} } \sim L^{\frac{2\Delta \q}{\nu}-d}
 \sum_{j=1}^{L^d}{j^{-\frac{2\Delta \q}{\nu d}}}
  .
\label{21}
\end{equation}
Matching both sides, and using the  static scaling relations 
$ 2\beta + \gamma  =  2 - \alpha$ and 
$ \beta (\delta - 1)   = \gamma$,  
one obtains
%\mycheck{\red{hyperhyperscaling}}
\begin{equation}
 \frac{\nu d}{\q} = 2 - \alpha .
\label{hyperhyperscaling}
\end{equation}
\index{scaling relation! hyperscaling}

This recovers  standard hyperscaling  (\ref{Jo})   provided $\q=1$ below $d_c$.
Above $d_c$, it delivers $\nu d_c = 2-\alpha$, which is also correct.

The  FSS $\chi_L \sim L^2$, supposed to hold above $d_c$ for FBC's, corresponds to  $\q=1$.
Eqs.(\ref{chih}) and (\ref{hjL}) would then deliver a  leading logarithm in the $d=6$ Ising case.
Such a result is spurious; leading logarithms can occur {\emph{only}} at the upper critical dimension \cite{Br82,KJJ2006,WeRi73,ourhighD}. 
Indeed, Butera and Pernici have recently given convincing evidence for the absence of leading logarithms in the Ising and scalar-field models using high-temperature series expansions in the thermodynamic limit in six dimensions \cite{Butera}.
This indicates that $\q \ne 1$ {\emph{even for FBC's}}. 

These conclusions are backed up by numerical simulations.
For the PBC case, however, either of the above paradigms supports the scaling forms $m_L \sim L^{-d/4}$ and $\chi_L \sim L^{d/2}$, so these cannot be used to discriminate between the two scaling scenarios. 
Direct numerical confirmation of Eq.(\ref{mainq}) in the PBC case, however, 
was given in Refs.[\refcite{JoYo05,ourNPB,ourCMP}].
The  formula (\ref{hjL}) for the Lee-Yang zeros was also verified for PBC's in Ref.~[\refcite{ourNPB}].
%Also in Ref.~[\refcite{ourNPB}] it was confirmed that the same FSS holds for PBC's at the pseudocritical point.

As discussed above, the long-standing belief is that Landau FSS rather than $Q$-FSS holds for FBC's above the upper critical dimension\cite{RuGa85,LuMa11,Watson,Gunton}.
%Indeed, recent, explicit numerical results purporting to support $\chi_L \sim L^{\gamma / \nu} = L^2$ at the critical point was given in Ref.~[\refcite{LuMa11}].
In Refs.~[\refcite{ourNPB,ourCMP}], however, numerical evidence {\emph{against}} the conventional  picture and {\emph{in favour}} of $Q$-FSS for FBC's was presented.
The claim is that $Q$-FSS applies in the FBC case above $d=4$, just as it does in the PBC case.
To observe it, one must perform FSS at the pseudocritical point for FBC's, not at the critical one.
The claim is that the infinite-volume critical point is so far from the pseudocritical point as to be outside the FSS window.

For a size-$L$ hypercubic FBC lattice  the  boundary sites interact with fewer than $2d$ nearest neighbours. 
In this sense,  they belong to a manifold of lower dimensionality.  
To truly probe the dimensionality of the FBC lattices,  the contributions of the outer layers of sites can be removed from the susceptibility and from other observables. 
To implement this in Refs.~[\refcite{ourNPB,ourCMP}], the contributions of the $L/4$ sites near each boundary were removed and only the contributions of the $(L/2)^d$ interior sites kept. 
For technical details of the simulation and determination of the core thermodynamic functions, the reader is referred to Refs.~[\refcite{ourNPB,ourCMP}].
The resulting FSS for the susceptibility  at pseudocriticality for $d=5$ is represented in Fig.~\ref{fig1}.

%...................................................................................
\begin{figure}[t]
\centerline{\includegraphics[width=0.55\textwidth]{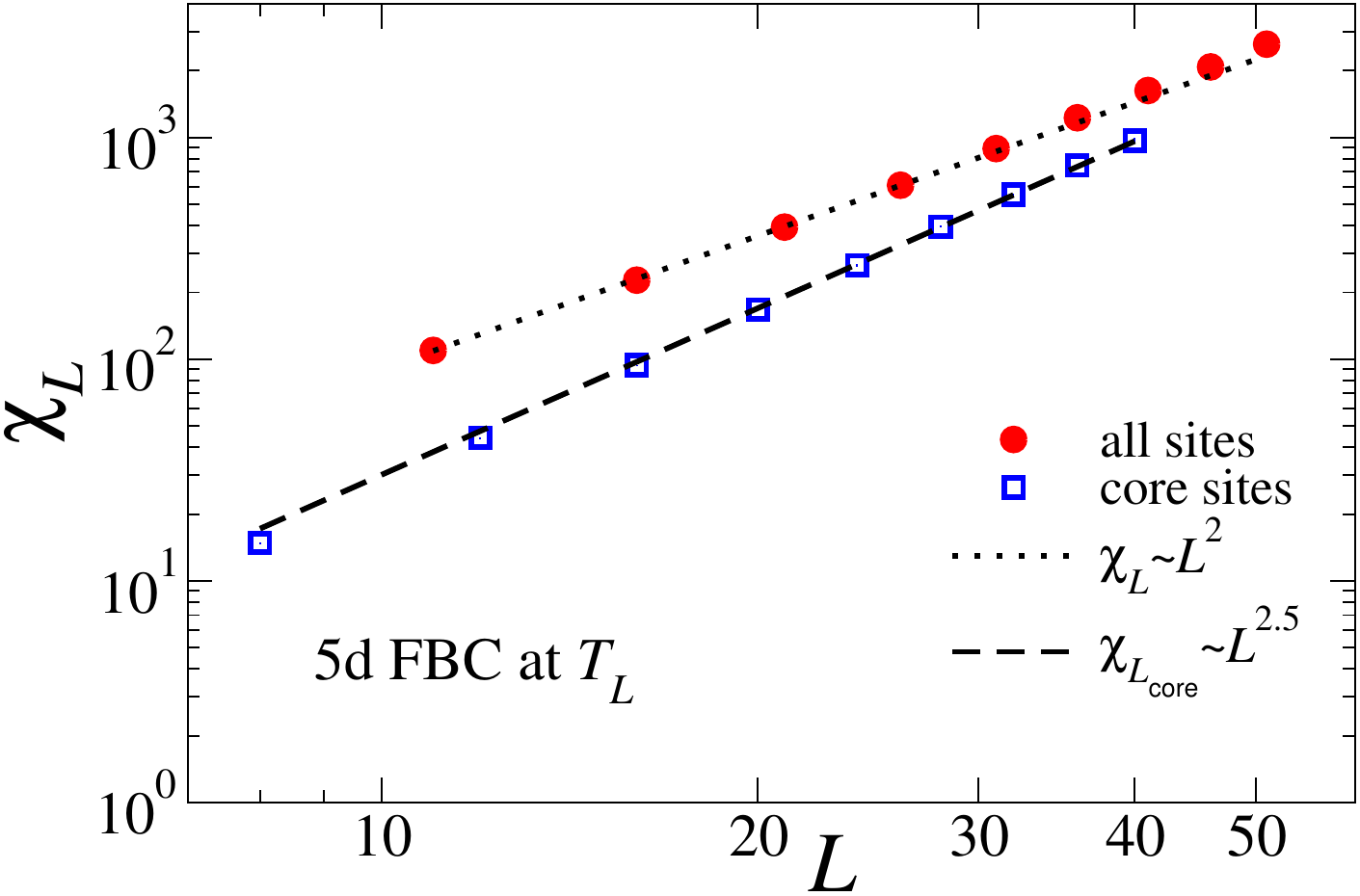}}
\caption{FSS for the 5D   Ising susceptibility at pseudocriticality.
For the upper data set (red) all sites are used in calculating $\chi_L$. 
For the lower set (blue) only core sites contribute. 
While the upper data set may appear to scale as $\chi_L \sim L^2$, in accordance with Gaussian FSS, this is spurious and due to surface sites. 
The scaling of the lower data, which are genuinely five-dimensional, supports the $Q$-FSS form $\chi_L \sim L^{5/2}$. } 
\label{fig1}
\end{figure}
%...................................................................................

The upper data set corresponds to using all sites in the calculation of the susceptibility.
The fit to the form $\chi_L \sim L^2$ (dotted line), suggested by the traditional Gaussian FSS paradigm for FBC lattices, appears reasonable at first sight.
However, closer inspection shows some deviation of the large-$L$ data from the line. 
We interpret this as signaling that the apparent fit to $\chi_L \sim L^2$ is spurious.
The lower data set corresponds  using only the interior $L/2$ lattice sites in the calculation of $\chi_L$. 
The best fit to the $Q$-FSS form  $\chi_L \sim L^{\q \gamma / \nu}  = L^{5/2}$ (dashed line) describes the large-$L$ data well.
This is evidence that the Ising model defined on the five-dimensional core of the  $L^5$ lattices obeys $Q$-FSS (\ref{II31k}) rather than Gaussian or Landau FSS. 
This can be interpreted as evidence for the universality of $\q$.

The susceptibility is closely related to the Lee-Yang zeros, as we have seen.
The FSS for the first two Lee-Yang zeros is presented in Fig.~\ref{fig2} at the pseudocritical point for FBC's using the contributions from all sites and from the core-lattice sites only.
In each case the zeros scale with as $L^{-\q\Delta / \nu} = L^{-15/4}$ according to $Q$-FSS.
There is no evidence to support the Gaussian prediction that $h_j(L) \sim L^{-\Delta / \nu} = L^{-3}$.

Thus we arrive at a new {\emph{universal}} scaling picture, to replace the old non-universal one.
The FSS ansatz is Eq.(\ref{II31k})  with correlation length given by Eq.(\ref{mainq}).
When $d>d_c$, $\q=d/d_c$ and when $d<d_c$,  $\q=1$ and the FSS ansatz reverts to Eq.(\ref{II31}).
There is no requirement for the thermodynamic length in the  scaling ansatz.
Also, hyperscaling is extended through Eq.(\ref{hyperhyperscaling}) beyond the upper critical dimension.
The new exponent $\q$ is therefore both {\emph{physical}} and {\emph{universal}} in the new picture.
Physically it controls the finite-size correlation length. 
We refer to this picture as $Q$-theory or $Q$-FSS, to distinguish it from Gaussian or Landau  FSS.

%...................................................................................
\begin{figure}[t]
\centerline{\includegraphics[width=0.55\textwidth]{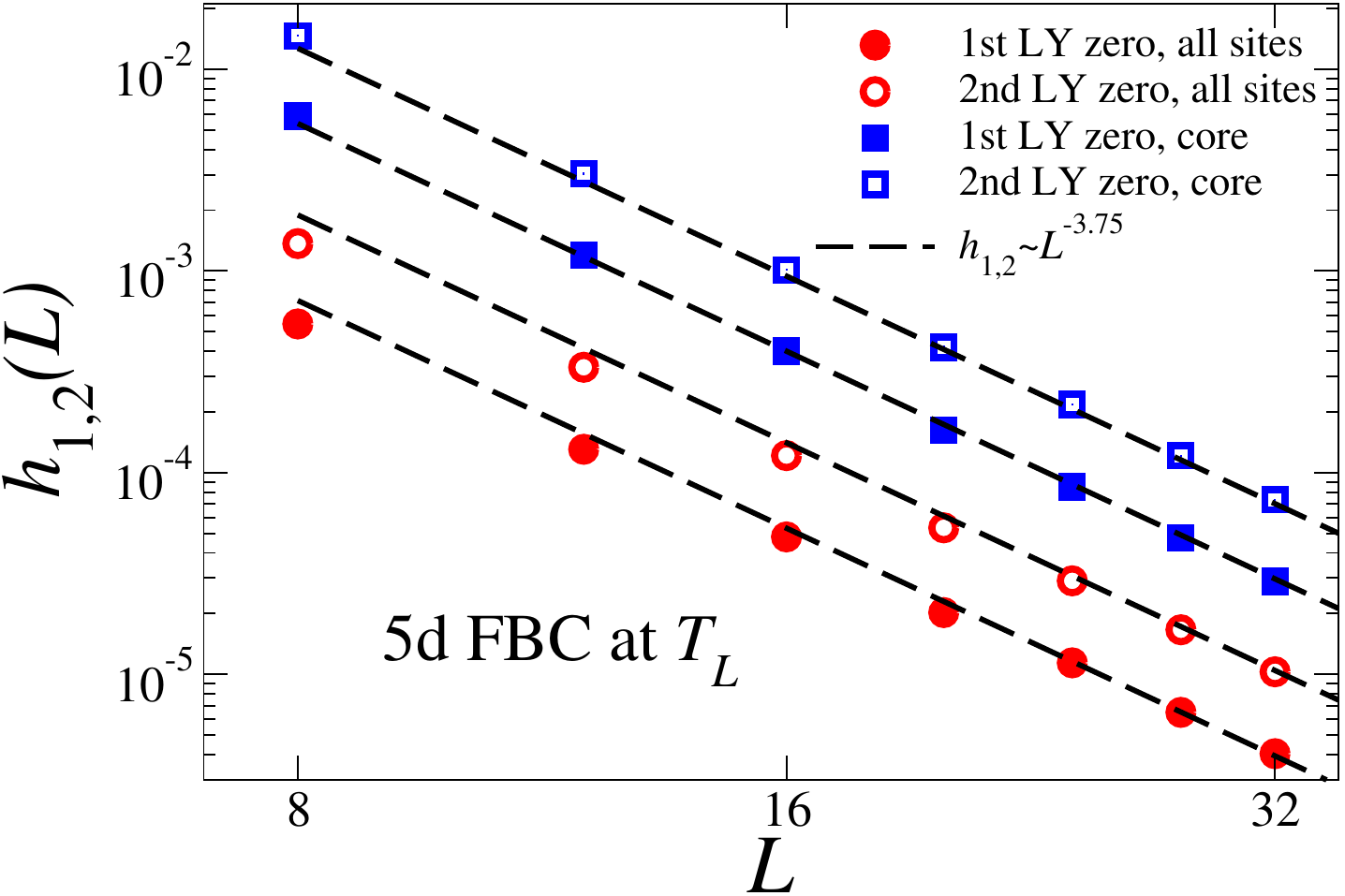}}
\caption{The first two Lee-Yang zeros for Ising systems with FBC's at pseudocriticality obey $Q$-FSS whether the full lattice or only the core sites are used in their determination. } 
\label{fig2}
\end{figure}
%...................................................................................

Thus we conclude that the numerical evidence is in favour of $Q$-FSS in the FBC case at pseudocriticality, just as it is for PBC's. 
It turns out that at the critical point itself, however, neither $Q$-FSS or Gaussian FSS are supported when one examines the core 5D lattices.\cite{ourNPB,ourCMP}
The reason for this is that $T_c$ lies outside the scaling window. 
The FSS window is measured by the rounding of the susceptibility peak.
In other words, the shifting $T_L-T_c$ exceeds the rounding.
The reader is refered to Ref.\refcite{ourNPB,ourCMP} where evidence supporting this interpretation is given.

This now brings us to re-examine the correlation function.

%%%%%%%%%%%%%%%%%%%%%%%%%%%%%%%%%%%%%%%%%%%%%%%%%%%%%%%%%%%%%%%%%%%
\subsection{Fisher's Scaling Relation and the Exponent $\eta_Q$}
\label{EPLF}
%%%%%%%%%%%%%%%%%%%%%%%%%%%%%%%%%%%%%%%%%%%%%%%%%%%%%%%%%%%%%%%%%%%

\index{scaling relation! Fisher}

In Ref.~[\refcite{ourEPL}] a new explanation for the negativity of the measured value of the anomalous dimension was proposed, based on the $Q$-theory outlined here\cite{ourNPB,ourCMP}.
According to the $Q$-FSS, there is a difference between the underlying length scale $L$ of the system above $d_c$ and its correlation length  $\xi_L$. 
In Eq.(\ref{G2}), the distance $r$ is implicitly measured on the correlation length scale and this leads to the usual Fisher relation (\ref{Fi}) via the process outlined in Section~\ref{11.05.2009c}.
In Sec.~\ref{fshortcomings}, however, the length scales $L$ and $\xi_L$ are incorrectly mixed
above the upper critical dimension.

To repair this, we firstly note that Eq.(\ref{mainq}) with $\q = d/d_c$ may be written
%\mycheck{\red{dims}}
\begin{equation}
 \xi_L^{d_c} = L^d.
\label{dims}
\end{equation}
Thus the correlation volume matches the actual volume when the correct dimensionalities are used, and that for $\xi_L$ is $d_c$ rather than $d$.

We write the critical correlation function in terms of the system-length scale  as
%\mycheck{\red{GscaleL}}
\begin{equation}
\ourG(0,r) \sim r^{-(d-2+\oureta)} \ourD \left({\frac{r}{L}}\right),
\label{GscaleL}
\end{equation}
where $\oureta$ is the anomalous dimension measured on this scale.
The subscript reminds that $Q$-FSS rather than standard FSS prevails there.

Integrating over space, then, the susceptibility is 
\begin{equation}
 \chi_L(0) \sim \int_0^L{\!\!r^{1-\oureta} \ourD \left({\frac{r}{L}}\right) dr} = 
 L^{2-\oureta} \!\!\! \int_{0}^{1}{{\!\!\ourD}(y) y^{1-\oureta}d y}.
\label{integral4}
\end{equation}
Above $d=d_c$, the $Q$-FSS formulae (\ref{FSSchi}) then yields
%\mycheck{\red{QFisher}}
\begin{equation}
  \oureta= 2 - \frac{\q \gamma }{ \nu}.
\label{QFisher}
\end{equation}
In the Ising case, for which $\q = d/4$, $\gamma = 1$ and $\nu = 1/2$, 
this gives $\eta_Q = 2-d/2$.
This is then identified with  $\eta^*$ of Refs.~[\refcite{NaBo70,BaGo73,LuBl97b,Luijtenthesis}]. 
The expression (\ref{integral4}) with $\eta_Q=-1/2$ in 5D is confirmed in Figs.~\ref{fig3} and~\ref{fig4}. 

%The new scaling relation (\ref{QFisher}) is the fluctuation-response theorem above the upper critical dimension when distance is measured on the scale of system size.
%The standard expression (\ref{Fi}) is the corresponding formula there   when distance is measured on the correlation-length scale.

%...........................................................................
\begin{figure}[t]
\begin{center}
\includegraphics[width=0.55\columnwidth, angle=0]{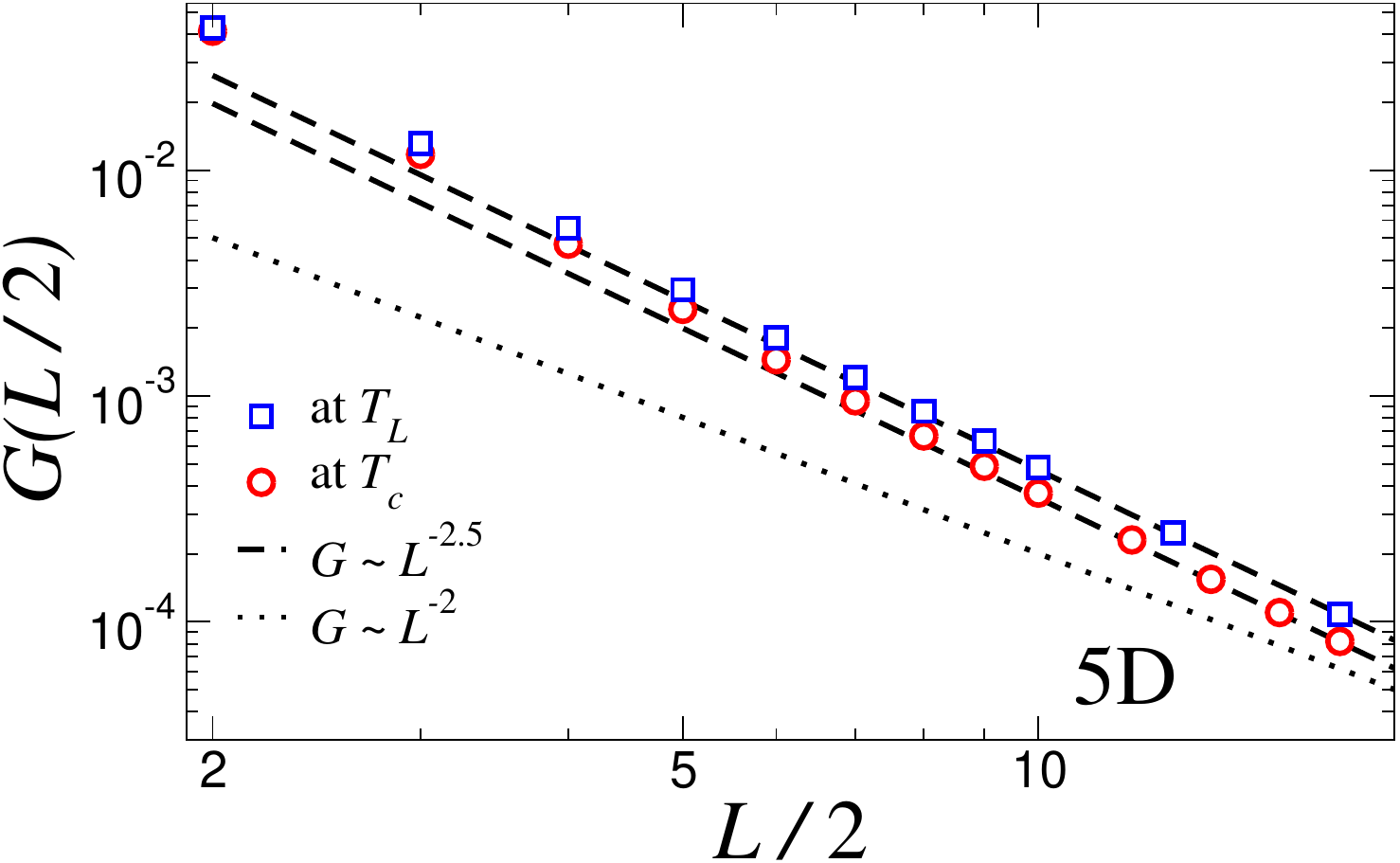}
\caption{The measured decay of the correlation function in five dimensions favours the $Q$-prediction $G_Q(t,L/2) \sim L^{-5/2}$ (dashed lines), corresponding to a negative value $\eta_Q=-1/2$, over the Landau prediction of a vanishing anomalous dimension (dotted line).}
\label{fig3}
\end{center}
\end{figure}
%...........................................................................

There are therefore two forms for the correlation function, two anomalous dimensions and two Fisher  relations, depending on whether distance is measured on the scale of $L$ or $\xi_L \sim L^{\qq}$.
The usual value $\eta = 0$ is correct only when distance is measured on the scale of correlation length. 
The value $\oureta = 2-d/2$ is correct on the length-scale $L$. 
Both are valid as characterising long-distance correlation decay.
In this interpretation, the notion of two different decay modes, one for short long distances and one for long long distances, is abandoned.
The new interpretation is fully consistent with explicit calculations. \cite{NaBo70,BaGo73,LuBl97b,Luijtenthesis,ourEPL}

The relationship between the two anomalous dimensions is 
\begin{equation}
 \oureta = \q \eta + 2(1-\q).
\label{etaetaQ}
\end{equation}
Below $d_c$, they coincide, while above  $d_c$, the $Q$-anomalous dimension $\oureta$ is negative.
Since the theorems on the non-negativity of  the $\phi^4$ anomalous dimension 
refer to correlation decay on the scale $\xi$, they refer to $\eta$ rather than $\oureta$, and are not violated~\cite{Fi64sc,Fi69inequalities,Delamotte}.

%...........................................................................
\begin{figure}[t]
\begin{center}
\includegraphics[width=0.55\columnwidth, angle=0]{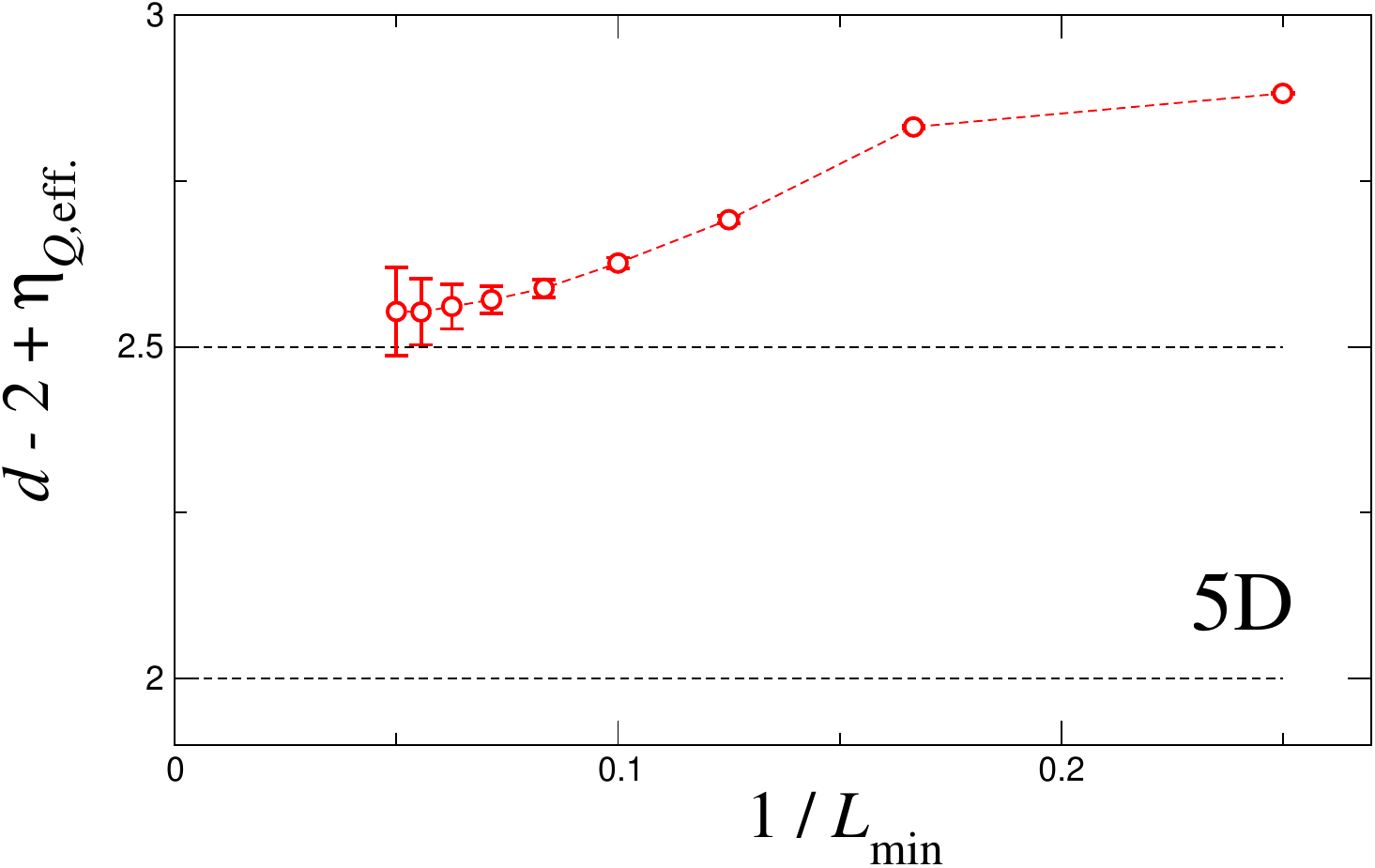}
\caption{As the minimum lattice size used in the fit increases, the measured  exponent in five dimensions approaches  the $Q$-prediction $\eta_Q=-1/2$. }
\label{fig4}
\end{center}
\end{figure}
%...........................................................................

%%%%%%%%%%%%%%%%%%%%%%%%%%%%%%%%%%%%%%%%%%%%%%%%%%%%%%%%%%%%%%%%%%%%%%%%%%s
%\mycheck{\red{fullthy}}
\section{The Full Scaling and FSS Theory above the Upper Critical Dimension}
\label{fullthy}
\setcounter{equation}{0}
%%%%%%%%%%%%%%%%%%%%%%%%%%%%%%%%%%%%%%%%%%%%%%%%%%%%%%%%%%%%%%%%%%%%%%%%%%%

In the previous section we confirmed $Q$-FSS numerically and demonstrated that it applies both to FBC's and PBC's.
We interpret this as evidence for universality of FSS and of the new exponent $\q$ above $d_c$.
We now need to revisit the renormalisation group scheme and show how the full scaling and FSS  theory arises through the mechanism of dangerous irrelevant variables.

%%%%%%%%%%%%%%%%%%%%%%%%%%%%%%%%%%%%%%%%%%%%%%%%%%%%%%%%%%%%%%%%%%%%%%%%%%s
\subsection{Leading Scaling Behaviour}
\label{fullQQQ}
%%%%%%%%%%%%%%%%%%%%%%%%%%%%%%%%%%%%%%%%%%%%%%%%%%%%%%%%%%%%%%%%%%%%%%%%%%%

We start with Eq.(\ref{BNPY1}) for the free energy:
%\mycheck{\red{BNPY11}}
\begin{equation}
 f_L(t,h,u) = b^{-d} f_{L/b}\left({ tb^{y_t}, hb^{y_h}, ub^{y_u}  }\right)
.
\label{BNPY11}
\end{equation}
Under the assumption of homogeneity (\ref{RG2}) we have the more compact form (\ref{RG3}).
Identifying $d^*$ with $d$ using the argumentation of Sec.~\ref{thermolength}, one has  $p_1=0$ in Eq.(\ref{RG4}).
The free energy is then
%\mycheck{\red{RG31}}
\begin{equation}
 f_L(t,h,u) = b^{-d} {\bar{f}}_{L/b}\left({ tL^{y_t^*}, hL^{y_h^*} }\right)
,
\label{RG31}
\end{equation}
in which  mean-field values $\alpha = 0$, $\beta = 1/2$, $\gamma = 1$ and $\delta = 1/3$ come from
%\mycheck{\red{BNPY91,BNPY101}}
\begin{eqnarray}
y_t^* & = & y_t - \frac{y_u}{2}  = \frac{d}{2}
, \label{BNPY91} 
\\
y_h^* & = & y_h - \frac{y_u}{4}  = \frac{3d}{4}
. \label{BNPY101} 
\end{eqnarray}
Appropriate differentiation of the free energy delivers the correct, Landau scaling in the thermodynamic limit and the correct $Q$-FSS forms (\ref{dfg2}), (\ref{dfg3}), irrespective of boundary conditions.

In Ref.~[\refcite{BNPY}], extensions of the above scenario to the correlation length or correlation function were not fully considered.
This is because (a) the finite-size correlation length was believed to be bounded by the system length and (b) the correct Landau values for  $\nu$ and $\eta$ were obtained by ignoring $u$.
Now we recognise that (a) is incorrect since $\q>1$ and (b) a new exponent $\eta_Q$ also emerges above $d_c$.

The equivalent form to Eq.(\ref{BNPY1}) for the correlation length is
\begin{equation}
 \xi_L(t,h,u) = b \xi_{L/b}\left({ tb^{y_t}, hb^{y_h}, ub^{y_u}  }\right)
.
\label{RG141}
\end{equation}
Following a similar procedure to that used for the free energy, the correlation length is expressed as a homogeneous function in Eq.(\ref{RG16}).
Eq.(\ref{RG17}) gives $\q = 1+q_1 y_u = d/4 $ provided $q_1 = -1/4$.
Setting $h=0$ and $L \rightarrow \infty$ recovers $\xi_\infty \sim t^{-\nu}$ provided $q_2 = -1/2 $ so that $y_t^{**} = d/2$.
Thus $y_t^{**}$ is the same as $y_t^*$ of Eq.(\ref{BNPY91}).
It is also the same as the scaling dimension required in Section~\ref{thermolength}, reaffirming that FSS is controlled by $\xi$ and the thermodynamic length $\ell$ is not required.
Moreover, $q_2 = p_2$ after Eq.(\ref{dfg11}).

To determine $q_3$ and $y_h^{**}$, we set  $t=0$ and let $L \rightarrow \infty$ and compare to $\xi_\infty(h,0)) \sim h^{-1/3}$ after Eqs.(\ref{xi}) and (\ref{X}).
One finds $q_3 = p_3 = -1/4$ and $y_h^{**} = y_h^* = 3d/4$.

Finally we return to the correlation function.
According to the standard paradigm, this is not affected by dangerous irrelevant variables. 
$Q$-theory, however, demands that it deliver correlation decay both on the scale of system length as well as on the correlation-length scale.

From dimensional analysis, we write
%\mycheck{\red{BNPYus}}
\begin{equation}
 G_L(t, u,r) = b^{-2d_\phi} G_{L/b} \left({tb^{y_t},ub^{y_u},rb^{-1}}\right),
\label{BNPYus}
\end{equation}
in which $d_\phi = d/2-1$.
Here,  $r$ and $b$ initially have the same dimension as $L$.
Acknowledging the danger of $u$, we treat this in a similar manner to the free energy and correlation length, and demand
\begin{equation}
 G_L(t,u,r) 
= b^{2-d+y_uv_1} \bar{G}_{L/b} \left({tb^{y_t^*}, rb^{-1+y_uv_2}}\right).
\label{BNPY3}
\end{equation}
Maintaining our interpretation of $r$ as a length requires $v_2=0$ so that the final argument on the right is dimensionless.
Setting $t=0$ and $b=r$ then, we obtain $G_L(0,u,r) = 1/r^{d-2-y_uv_1}$.
This agrees with $\ourG$ in Eq.(\ref{GscaleL}) provided that $v_1 = -\oureta/y_u = -1/2$.

If, on the other hand, we wish to interpret $r$ as a correlation length, the final argument is dimensionless if it is $rb^{-\q}$. 
We then require $v_2 = (1-q)/y_u = 1/4$.
Again setting $t=0$, but now with  $b^{\qq}=r$, we obtain 
 $G_L(0,u,r) = 1/r^{(d-2-y_uv_1)/\qq}$.
Inserting $v_1=-1/2$ delivers the Ornstein-Zernike\cite{OZ} form $G(0,u,r) \sim 1/r^2$, corresponding to the Landau theory.

%%%%%%%%%%%%%%%%%%%%%%%%%%%%%%%%%%%%%%%%%%%%%%%%%%%%%%%%%%%%%%%%%%%%%%%%%%%
\subsection{Logarithmic corrections}
\label{logsf}
%%%%%%%%%%%%%%%%%%%%%%%%%%%%%%%%%%%%%%%%%%%%%%%%%%%%%%%%%%%%%%%%%%%%%%%%%%%

\index{scaling! logarithmic corrections}

%...........................................................................
\begin{figure}[t]
\begin{center}
\includegraphics[width=0.55\columnwidth, angle=0]{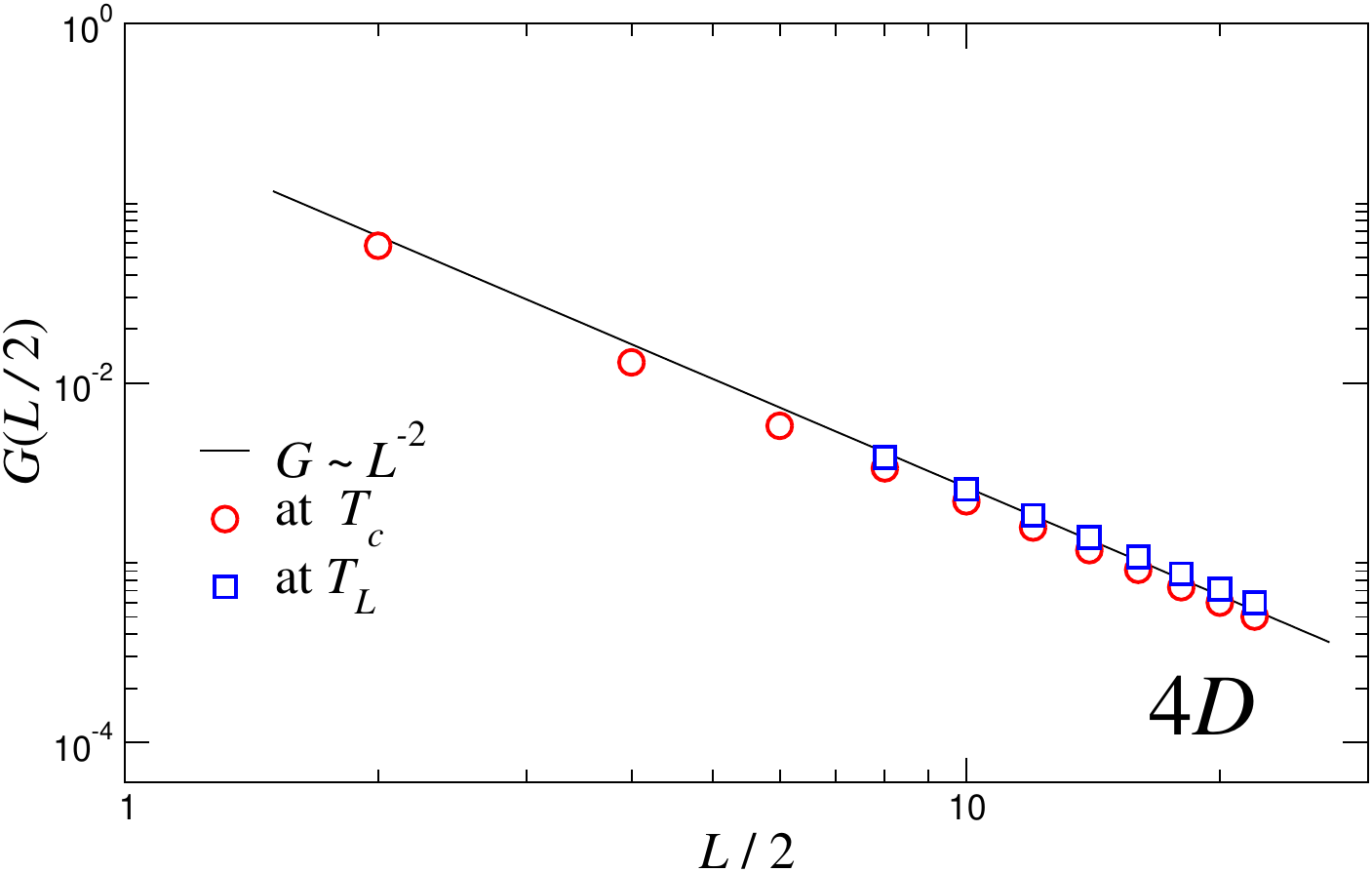}
\caption{The leading scaling for the correlation function in four dimensions has exponent $d_c-2+\eta = 2$, corresponding to vanishing anomalous dimension. This is predicted by both paradigms.}
\label{fig5}
\end{center}
\end{figure}
%...........................................................................

Although the theories underlying $Q$-scaling and the conventional paradigm differ at a fundamental level, each scheme predicts the same numerical values for the various critical exponents in the PBC case.
To decide between them, we examine the upper critical dimension itself.
The quartic variable $u$ is marginal there and  the renormalization-group formalism gives rise to logarithmic corrections. 
With no dangerous irrelevant variables, the old paradigm has no consequence at $d=d_c$: there is only one correlation function for long distances. 
In fact, taking logarithmic corrections into account, one expects at criticality
\begin{equation}
   G_{\xi}(0,r)   \sim  D\left({\frac{r}{\xi_L}}\right)
	 r^{-(d_c-2+\eta)} (\ln{r})^{\hat{\eta}} .
   \label{GP}
\end{equation}
The scaling relations for logarithmic corrections connect the logarithmic analogue of the anomalous dimension, $\hat{\eta}$ to $\eta$, $\hat{\gamma}$ and $\hat{\nu}$ through Eq.(\ref{SRlog4}), which was proposed and confirmed in the thermodynamic limit in Ref.~[\refcite{KJJ2006}] and  reported upon in the previous volume of this series \cite{KeVol3}.

$Q$-scaling theory, however, again leads to a new prediction here.
The $Q$-correlation function is
\begin{equation}
   G_L(0,r)   \sim  r^{-(d-2+\eta_Q)} (\ln{r})^{\hat{\eta}_Q} .
   \label{GQ}
\end{equation}
Of course, $\eta_Q=\eta $ at $d=d_c$.

The fluctuation-dissipation  theorem, with an appropriate bound $L$ for Eq.(\ref{GQ}) then gives\cite{ourEPL}
\begin{equation}
 \chi_L(0)  \sim  L^{2-\oureta}(\ln{L})^\ouretahat
 \left[{1+ {\mathcal{O}}} (1/\ln{L})\right],
\end{equation}
where
\begin{equation}
 \hat{\gamma}  =  (2 - \oureta) (\hat{\nu}-\hat{\q}) + \ouretahat.
 \label{61}
\end{equation} 

The two anomalous dimensions are related by
\begin{equation}
 \ouretahat = \hat{\eta} + (2-\eta)  \hat{\q} .
\label{etaqhat1o2}
\end{equation}

These new predictions are not derivable from the conventional paradigm and can be tested  numerically.\cite{ourEPL}
In Fig.~\ref{fig5},  $G(t,L/2)$ is plotted against  $L/2$ at both the critical and pseudocritical points for periodic lattices, confirming the leading behaviour is governed by a vanishing anomalous dimension.
The $Q$-prediction is that $\ouretahat =1/2$ in the $d=4$ Ising model.
This is confirmed in Fig.~\ref{fig6}, in which $(L/2)^2 G(t,L/2)$ is plotted against  $\ln{(L/2)}$.
The positive slope is clearly not $\hat{\eta}=0$ and compatibility with $\ouretahat=1/2$ is evident. 
%are nicely compatible with the numerical data.

%...........................................................................
\begin{figure}[t]
\begin{center}
\includegraphics[width=0.55\columnwidth, angle=0]{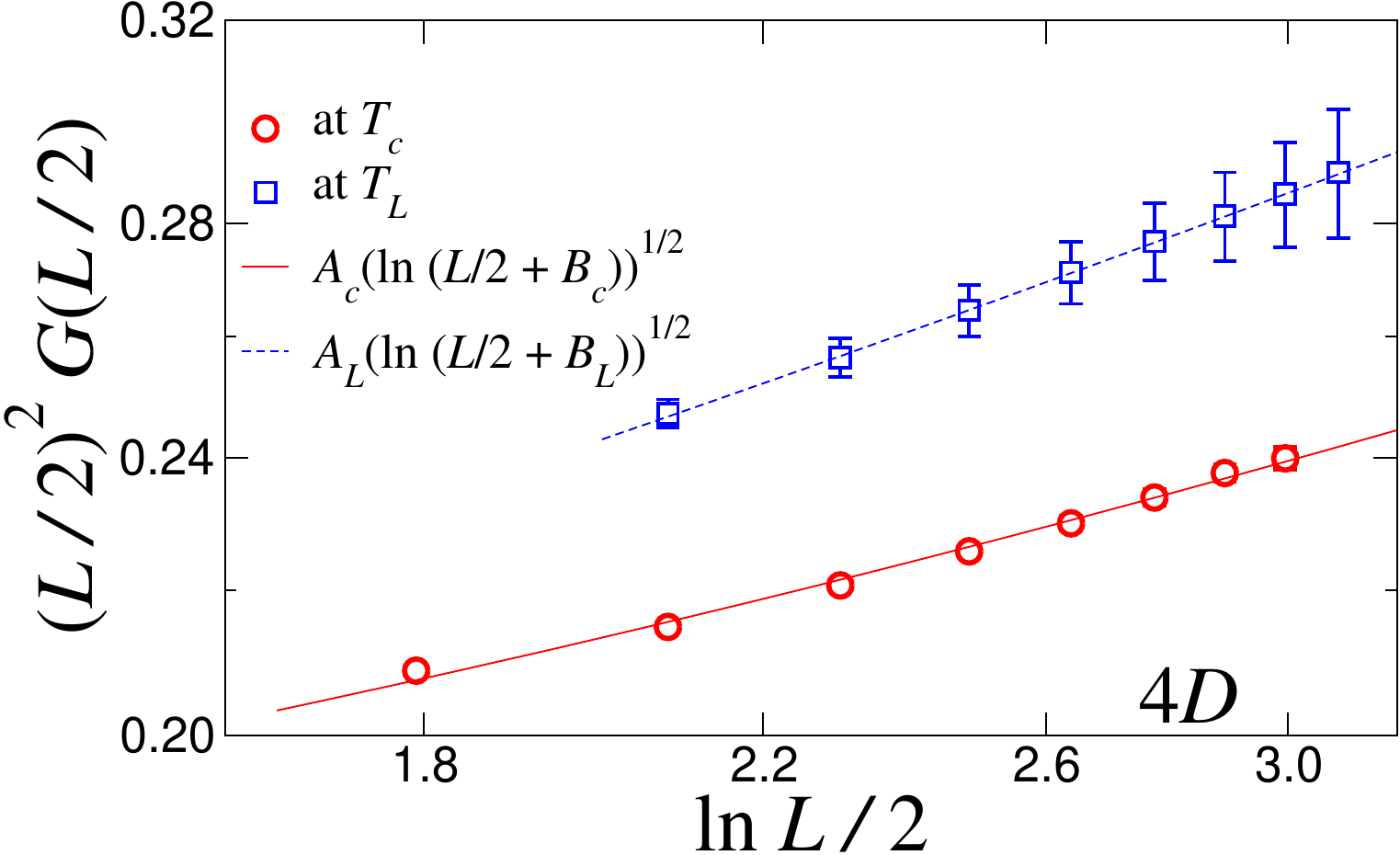}
\caption{The logarithmic correction to scaling for the measured correlation function has a positive exponent compatible with the $Q$-prediction $\hat{\eta}_Q=1/2$ and incompatible with the Landau value zero. }
\label{fig6}
\end{center}
\end{figure}
%...........................................................................

%%%%%%%%%%%%%%%%%%%%%%%%%%%%%%%%%%%%%%%%%%%%%%%%%%%%%%%%%%%%%%%%%%%%%%%%%%%
%\mycheck{\red{Conclusions}}
\section{Conclusions}
\label{Conclusions}
\setcounter{equation}{0}
%%%%%%%%%%%%%%%%%%%%%%%%%%%%%%%%%%%%%%%%%%%%%%%%%%%%%%%%%%%%%%%%%%%%%%%%%%%

Mean-field and Landau theories are presented in the early chapters of many textbooks on critical phenomena  because (a) they are a simple theories which manifest phase transitions and (b) they form the starting point for the development of more sophisticated and realistic theories. 
Despite their simplicity, the Ginzburg criterion indicated that mean-field and Landau theories should  deliver a quite accurate and full account of scaling above the upper critical dimension. 
In particular, they yield the  critical exponents $\alpha = 0$, $\beta = 1/2$, $\gamma = 1$, $\delta = 1/3$, $\nu = 1/2$ and $\eta = 0$.
These satisfy all scaling relations except  hyperscaling (\ref{Jo}) when $d>4$.

On the theoretical side, it has long  been recognised that, to match the renormalizaton group to the predictions of mean-field and Landau theories in the thermodynamic limit, one needs to take into account dangerous irrelevant variables above the upper critical dimension. While this delivers the correct values of the critical exponents, a naive application of FSS then delivers $m_L \sim L^{-1}$ and $\chi_L \sim L^2$, for example.

These are known to be incorrect for systems with PBCs where, instead, $m_L \sim L^{-d/4}$ and $\chi_L \sim L^{d/2}$.
To remedy this, a new thermodynamic length scale was introduced to control FSS above $d_c$.
Although this is an infinite-volume concept, it was supposed only to affect FSS in the case of periodic boundaries; Gaussian FSS was supposed to apply when the boundaries are free. Thus FSS loses its universality above the upper critical dimension.

Numerical simulations also indicated a puzzle related to the correlation function.
Despite theorems to the contrary, numerical studies indicated a negative value for the anomalous dimension, while Landau theory predicts that $\eta$ vanishes.
To explain this, the idea was formed that there is a difference between short long range order and long long range order. 
Thus the standard paradigm appears to involve  ad hoc fixes to a number of puzzles. 

Here we have reported on recent advances to scaling theory above the upper critical dimension.
Firstly it is recognised that earlier bounds on the scaling of the correlation length are inapplicable.
In fact numerical evidence supports non-trivial FSS for $\xi_L$ and an analytical argument requires the introduction of a new exponent (denoted $\q$) to track it.
The same numerical argument suggests that the new exponent $\q$ is universal as well as physical and this is supported by careful FSS for systems with free boundaries.

These considerations also prompted the re-examination of the correlation function above the upper critical dimension. The claim is that its functional form depends upon whether distance is measured on the system length scale or on the scale of the correlation length. In the latter case, applicable in the thermodynamic limit, the anomalous dimension vanishes and Fisher's scaling relation applies.
However, on the scale of the system length, a new  anomalous dimension $\eta_Q$ has to be introduced. In the case of the Ising model, this is negative and brings with it a new Fisher-type scaling relation.
Under the new scaling paradigm, there is no thermodynamic length and no difference between short long distances and long long distances. These were ad hoc features of the old scaling picture.

The new scaling theory also predicts new results at the upper critical dimension itself. 
This is the case despite the absence of dangerous irrelevant variables there.
In particular, new logarithmic corrections to the correlation function are predicted and tested.
The results support the new $Q$-scaling theory.

Revisiting the Ginzburg criterion, in Sec.~\ref{secGinzburg}, we see that the replacement of the finite-size volume $N$ by $\xi_\infty^d$ leading to the thermodynamic-limit relation (\ref{Ginz3}) is unjustified.
That expression should rather read $ \chi_\infty/\beta \ll \xi_\infty^{d_c} m_\infty^2$.
The $Q$-corrected Ginzburg criterion then reads $\nu d_c >  2-\alpha$, which is not satisfied as an inequality.
This means the Ginzburg criterion is invalid, even above $d_c$, so that strictly speaking, mean-field theory is not a full description of scaling there.

%%%%%%%%%%%%%%%%%%%%%%%%%%%%%%%%%%%%%%%%%%%%%%%%%%%%%%%%%%%%%%%%%%%%%%%%%%%
%\section*{Acknowledgments}
%%%%%%%%%%%%%%%%%%%%%%%%%%%%%%%%%%%%%%%%%%%%%%%%%%%%%%%%%%%%%%%%%%%%%%%%%%%

%Acknowledgments to funding bodies etc. may be placed in a separate
%section at the end of the text, before the Appendices. This should not
%be numbered so use \verb|\section*{Acknowledgements}|.

%%%%%%%%%%%%%%%%%%%%%%%%%%%%%%%%%%%%%%%%%%%%%%%%%%%%%%%%%%%%%%%%%%%%%

%%%%%%%%%%%%%%%%%%%%%%%%%%%%%%%%%%%%%%%%%%%%%%%%%%%%%%%%%%%%%%%%%%%%%

%%%%%%%%%%%%%%%%%%%%%%%%%%%%%%%%%%%%%%%%%%%%%%%%%%%%%%%%%%%%%%%%%%%%%
%\printindex[aindx]                 % to print author index
\printindex                         % to print subject index
%%%%%%%%%%%%%%%%%%%%%%%%%%%%%%%%%%%%%%%%%%%%%%%%%%%%%%%%%%%%%%%%%%%%%

\end{document}